# Near-optimal quantization and linear network coding for relay networks

Anand Muralidhar and P. R. Kumar

**Abstract**

We introduce a *discrete network* corresponding to any Gaussian wireless network that is obtained by simply quantizing the received signals and restricting the transmitted signals to a finite precision. Since signals in the discrete network are obtained from those of a Gaussian network, the Gaussian network can be operated on the quantization-based digital interface defined by the discrete network. We prove that this digital interface is near-optimal for Gaussian relay networks and the capacities of the Gaussian and the discrete networks are within a bounded gap of $O(M^2)$ bits, where $M$ is the number of nodes.

We also prove that any near-optimal coding strategy for the discrete network can be naturally transformed into a near-optimal coding strategy for the Gaussian network merely by quantization. We exploit this property by designing a linear coding strategy for the case of layered discrete relay networks. The linear coding strategy is near-optimal and achieves all rates within $O(M^2)$ bits of the capacity, independent of channel gains or SNR. The linear code is therefore a near-optimal strategy for layered Gaussian relay networks and can be used as-is on the Gaussian network after simply quantizing the signals. The linear code is also robust and the relays need not know the channel gains on either the incoming or the outgoing links. The transmit and receive signals at all relays are simply quantized to binary tuples of the same length $n$, which is all that the nodes need to know. The linear network code is a particularly simple scheme and requires all the relay nodes to collect the received binary tuples into a long binary vector and apply a linear transformation on the long vector. The resulting binary vector is split into smaller binary tuples for transmission by the relays. The quantization requirements of the linear network code are completely defined by the parameter $n$, which therefore also determines the resolution of the analog-to-digital and digital-to-analog convertors that are required for operating

Anand Muralidhar is with Alcatel-Lucent Bell Labs, India, and P. R. Kumar is with the Dept. of Electrical and Computer Engineering at Texas A&M University, College Station, USA: {anand.muralidhar@alcatel-lucent.com, prk@tamu.edu}.

This material is based upon work partially supported by NSF under contracts CNS-0905397 and CNS-1035340, AFOSR under Contract FA9550-09-0121, and and USARO under Contract W911NF-08-1-0238.

the network within a bounded gap of the network's capacity. As is evident from the description, the linear network code explicitly connects network coding for wireline networks with codes for Gaussian networks.

# I. Introduction

Even in a simple wireless network, there is an infinitude of options and techniques available for cooperation among nodes, and it is of great interest to determine provably near-optimal operating strategies. This has motivated research into network information theory that aims to determine the fundamental limits on the capacity of wireless networks as well as optimal or near-optimal strategies for operating them [21]. This is generally a difficult problem and a precise characterization has remained elusive for most networks. One approach to circumvent this is to obtain scaling laws that shed light on capacity as well as operating strategies, as the number of nodes in a network increases [8]. Another approach, initiated in the pioneering work of [2], is to seek answers that are within a bounded gap of the capacity that is independent of the channel gains or SNR, and is a function of only the number of nodes in the network. This is done by studying an alternative model that is more tractable then the original Gaussian network. Such a bounded gap approximation is valid at all SNRs, and is most relevant in the high power or low noise regimes.

In this paper we focus on the latter direction. We address the issues of both approximating the capacity of Gaussian relay networks, and rigorously prove the near-optimality of a coding strategy to operate them. Our approach employs the strategy espoused in [2] – study not the original Gaussian network, but an alternative model. Our goal in this paper is to develop just such an alternative model that can serve as a digital interface for operating Gaussian networks in that it simultaneously possesses three properties. First, the alternative model must well-approximate the capacity of the Gaussian network up to a constant gap; i.e., the capacity of the original Gaussian network and the capacity of the alternative model must be within a bounded gap that is independent of channel gains or SNR. Second, it must be possible to easily operate the Gaussian network on the interface defined by the new model. The first two properties give us the capability to convert any near-optimal coding strategy for the alternative model into a near-optimal coding strategy for the original Gaussian network. Their usefulness is however contingent on actually being able to design a near-optimal strategy for the alternative model.

So, the third property is that the alternative model be tractable in that one can actually design near-optimal strategies for it.

In an earlier paper [1] a superposition model was proposed that had the first property. However, Gaussian networks cannot be naturally operated on the interface defined by the model, and near-optimal coding strategies for the superposition model are not easy to design or describe. The truncated deterministic model proposed in [2] satisfies the first property but has drawbacks similar to that of the superposition model. The linear deterministic model proposed in [2] has the third property since optimal codes can be designed for it, but not the first property since it cannot approximate the capacity of the Gaussian network. It satisfies the second property in a limited context since a strategy in the linear deterministic network need not get mapped to an identical strategy in the Gaussian network. In this paper, we propose a different model, called a *discrete model*, that has all three properties for the class of layered Gaussian relay networks. In fact, for this model we show that a linear network code is actually near-optimal. Thus we establish the interesting result that scalar quantization followed by linear network coding is near-optimal for the class of Gaussian layered relay networks, achieving rates within a bounded gap of capacity.

The discrete model is naturally derived from the Gaussian model by simply quantizing the received signals and by restricting the transmit signals to a finite alphabet. Thus the stochasticity of the Gaussian network carries through into the discrete model and we can construct a discrete network that corresponds to any Gaussian network. Note that the discrete model is not a noiseless deterministic model as in [2] or [1]. In order to operate the Gaussian network on the discrete interface, one simply quantizes the received signals and restricts the transmit signals to a finite alphabet. Hence the discrete network provides a *quantization-based digital interface* for operating Gaussian networks. We prove that the discrete network satisfies the above-mentioned three properties for the class of layered Gaussian relay networks. The capacities of the layered Gaussian relay network and its discrete counterpart are shown to be within a bounded gap of $O(M^2)$ bits, where $M$ is the number of nodes in the network. We construct a simple coding scheme for the layered discrete relay network, called the *linear network code*, which achieves all rates within $O(M^2)$ bits of the capacity of the discrete network, independently of channel gains or SNR. This linear network code is also near-optimal for the layered Gaussian network and can be used on it by operating the Gaussian network on the discrete interface. The linear network code is a generalization of random linear network coding for wireline networks [5], [6], or for linear deterministic networks [2]. As explained in detail in the sequel, the linear network code involves random encoding at the source followed by a random linear transformation at the relay. The crucial distinction between this scheme and the network coding schemes in the literature is that we need to account for the noise in the received signals at the relays and destination. To the best of our knowledge, this is the first instance of an explicit connection between network coding for wireline and noiseless networks with coding for Gaussian networks, with a bounded-gap guarantee on performance.

### A. *Previous work on relay networks*

Any summary of work on relay networks has to start with the early and lasting contributions in [4] and [3]. Networks with a single relay node were introduced in [3], and the well-known *decode-and-forward* (DF) and *compress-and-forward* (CF) schemes for such relay networks were developed in [4]. DF was extended to networks with multiple relays in [9], [10]. An extension of CF to multiple relays was presented in [11]. It was proved in [2] that these extensions of DF and CF to relay networks achieve rates which can be arbitrarily lower than the network capacity.

In [2], a fundamental question on the tightness of the cut-set upper bound on the capacity of a relay network was answered in the affirmative. It was proved that the capacity of the relay network is at most $O(M \log M)$ bits below the cut-set bound, where $M$ is the total number of nodes in the network, and the bounded gap is independent of channel gains or the SNR. The result was established via constructing a new coding scheme called *quantize-map-and-forward* (QMF) in which the relays quantize the received signals, buffer them, and randomly map them to a Gaussian codeword. QMF achieves all rates within $O(M \log M)$ bits of the cut-set bound.

Another important contribution in [2] was the introduction of the *linear deterministic model* as a technique to approximate Gaussian networks and to gain insights into coding schemes for Gaussian networks. For every Gaussian network, we can construct a corresponding linear deterministic network. The cut-set bound is the capacity of the linear deterministic relay network and can be achieved by random linear codes. However it was shown in [2], [1] that the linear deterministic model does not approximate the capacity of Gaussian relay networks and the capacity of the Gaussian network can be arbitrarily higher than that of the corresponding linear deterministic network. It was proved in [2] that the *truncated deterministic model* can approximate the capacity of Gaussian relay networks within a gap of $O(M \log M)$ bits. The *discrete superposition model* for networks was introduced in [1] as an alternate deterministic model for approximating Gaussian networks. It was shown that the capacity of the Gaussian and the corresponding discrete superposition networks are within $O(M \log M)$ bits. More importantly, it was proved in [1] that any coding scheme for the discrete superposition network can be simply lifted to the Gaussian network with a loss of at most $O(M \log M)$ bits in the rate. This reduced the problem of code design for Gaussian networks into the perhaps simpler problem of code design for the noiseless discrete superposition network, though the lifting procedure still had exponential complexity. However, the issue of designing a near-optimal code for the noiseless discrete superposition network remained elusive. The study of simple coding schemes for the discrete superposition network led to the linear network code presented here.

The bounded gap approximation for capacity of relay networks was further improved in [12] and an extension of compress-and-forward to relay networks was constructed that was approximately optimal. Here the relays are required to perform vector quantization in order to compress their received signals. In [13], the minimal compression rates for the relay nodes were computed, and the decoding procedure from [12] was further simplified.

The capacity results for relay networks have spurred research in finding low complexity coding schemes. In [14], the QMF scheme was modified by choosing low-density parity-check codes for encoding at the source and the relay instead of Gaussian codes. A simplified decoding algorithm based on Tanner graphs was presented and the viability of the proposed technique was shown via simulations. In [15], a different approach was taken by constructing codes that are computationally tractable when compared to QMF. A concatenated code was presented for the relay network where the outer code is a polar code and the inner code is a modification of the random Gaussian code from [2]. This approach was shown to have computational complexity that is near-linear in the block-length of the code. In [16], the quantization and encoding in QMF is modified by using nested lattice codes at the source and relays.

### B. Network coding

Network coding was introduced in the landmark paper [5]. The max-flow min-cut theorem was established for noiseless wireline networks with a single source and multiple destinations, which essentially implied that the cut-set bound was the capacity of these networks. A class of codes called $\beta$-codes, which involved random encoding at the source and intermediate nodes, were shown to achieve the capacity of the network with increasing block-length. There were other important contributions in [5]: namely, the notion of distinguishability of codewords at the destination, which naturally introduces the notion of cuts in proving the achievability of the cut-set bound, and the technique of time-parameterizing cyclic networks to view them as special cases of acyclic networks. Both these techniques were used in the proofs in [2] and are used in the proof of approximate optimality of linear network codes for Gaussian networks in this paper.

Later, in [17] and [18], it was established that linear network codes suffice to achieve the capacity of wireline networks. In [17], algebraic tools were introduced that simplified the analysis and design of network codes. In [6], it was shown that random encoding at the source and random linear operations at the relay nodes achieves the capacity of the wireline network in the limit of increasing block-length of the codewords. In [2], the applicability of random linear coding to linear deterministic networks was shown, where random encoding by the source and random linear encoding by the relays achieves the cut-set bound in the limit as the block-length tends to infinity. One of the contributions in this paper is to show the applicability and near-optimality of random linear coding for layered Gaussian networks.

### *C. Outline of the paper*

In Section II, we describe the network models. We define the Gaussian model for networks and also the discrete model, where the latter is obtained by quantizing the received signals in the Gaussian model and by restricting the transmit signals to a finite alphabet. Next we prove that the cut-set upper bounds on the capacities of the Gaussian and discrete networks differ in a bounded gap of $O(M \log M)$ bits, with the bound independent of channel gains or SNR (Section III). The linear network code for layered discrete networks is presented in detail in Sec. IV. The linear network code is shown to be approximately optimal and achieves rates within $O(M^2)$ bits of the capacity of the layered discrete network. In Sec. V, the linear network code is proved to be approximately optimal for the layered Gaussian network and achieves all rates within a bounded gap of $O(M^2)$ bits from the capacity of the network, with the bound independent of channel gains or SNR. This will imply that the capacities of the layered Gaussian network and the layered discrete network are within a bounded gap of $O(M^2)$ bits. The linear network code is also approximately optimal for layered MIMO networks where the nodes have multiple transmit and receive antennas (Sec. V-B), and for multicast networks where the source wants to transmit the same information to a subset of the other wireless nodes (Sec. V-C). We conclude by summarizing the properties of the linear network code in Sec. VI.

## II. Network models

First we describe the Gaussian model for wireless networks, followed by the discrete network that serves as the digital interface for the linear network code.

### A. Gaussian model

We consider relay networks consisting of a single source and a single destination, which we model by the following Gaussian network. The network has a total of $M+1$ nodes labeled as $\{0,1,\ldots,M\}$, with $0$ being the source and $M$ the destination. A directed graph describes the underlying topology of the network. The wireless link connecting two nodes $i$ and $j$ is described by a complex channel gain $h_{ij}$. All the transmit signals $\{x_i\}$ are constrained to be complex numbers with a unit power constraint. The received signal at node $j$ is given by

$$y_j = \sum_{i \in \mathcal{N}(j)} h_{ij} x_i + z_j, \tag{1}$$

where $\mathcal{N}(j) = \{i : (i,j) \in \mathcal{E}\}$ is the set of neighbors of the $j$-th node, $z_j$ is $\mathcal{CN}(0,1)$ complex white Gaussian noise independent of the transmit signals, and $h_{ij} = h_{ijR} + \imath h_{ijI}$.

We will also allow for MIMO relay networks where the nodes have multiple transmit and receive antennas. Suppose node $i$ has $T_i$ transmit antennas and $U_i$ receive antennas. In that case, the transmitted signal at each time instant at node $i$ is a $T_i$-dimensional vector of complex numbers, and the received vector is $U_i$-dimensional. Correspondingly, the channel between two nodes $i$ and $j$ is then described by a collection of channel gains $\{h_{ij}^{kl}\}$ where $k$ and $l$ index respectively the transmit antennas at node $i$ and the receive antennas at node $j$. The received signal at node $j$ is

$$y_j^l = \sum_{i \in \mathcal{N}(j)} \sum_{k=1}^{T_i} h_{ij}^{kl} x_i^k + z_j^l, \; l = 1, \ldots, U_j, \tag{2}$$

where $x_i^k$ is the signal transmitted from the $k$-th antenna of node $i$, and $y_j^l$ is the received signal at the $l$-th antenna of node $j$. $z_j^l$ is complex Gaussian noise added to the received signal at the $j$-th node. For simplicity of exposition, we restrict a transmit signal to satisfy an individual power constraint, though our results can be extended to allow for a total power constraint across all the antennas of a node.

### *B. Discrete model*

Now we describe a *quantization-based digital interface* for operating the above Gaussian network, i.e., for the purpose of defining the coding strategy. It is obtained, as the name above suggests, by quantizing the continuum valued signals received from the Gaussian network, and by restricting the choice of transmitted signals to lie in a discrete set. The quantization of the received signals is natural, though we will even discard sign information. We will call the overall network resulting from the discrete-inputs and discrete-outputs at each node as the *discrete network*.

In the language of automata theory, this discrete network can be *simulated* (see [19]) from the Gaussian network, i.e., it uses lesser information than the original Gaussian network. Our key result consists of three dovetailed parts. The first part is that there is no significant loss in the cut-set bound, in that the bit gap between the upper bounds on the capacities of the two networks, the Gaussian network and the discrete network, is bounded over all SNRs. The second result, described in Section IV, builds on this first key result. The second part is that we identify a near capacity achieving coding strategy for the discrete network. The third part is that this coding strategy for the discrete network can be easily implemented on the Gaussian network, through simple quantization, and achieves rates within a bounded bit gap from the capacity. Thereby we obtain both a capacity approximating discrete network, a simple linear coding strategy for the discrete network, as well as a natural mapping of the coding strategy from the discrete network to the Gaussian network. Together, the combined results establish a discrete network for both analyzing the original network, as well as operating it near-optimally through simple quantization. It is therefore appropriate to call such a discrete network with these two rigorously established properties as a *quantization-based digital interface.*

The *discrete network* is obtained by quantizing the received signals and constraining the transmit signals in a Gaussian network. The received and transmit signals are allowed to take finitely many values lying in what can essentially be regarded as a quadrature amplitude modulation (QAM) constellation. Define $n$ to be

$$n := \max_{(i,j) \in \mathcal{E}} \max\{\lfloor \log |h_{ijR}| \rfloor, \lfloor \log |h_{ijI}| \rfloor\}. \tag{3}$$

The channel inputs (transmit signals) in the discrete network are complex valued, with both real and imaginary parts taking values from $2^n$ equally spaced discrete points. The transmit symbol is

$$x = \frac{1}{\sqrt{2}}(x_R + \imath x_I), \tag{4}$$

where

$$x_R = \sum_{k=1}^{n} 2^{-k} x_R(k), \tag{5}$$

$$x_I = \sum_{k=1}^{n} 2^{-k} x_I(k), \tag{6}$$

with each $x_R(i)$ and $x_I(j)$ in $\mathbb{F}_2$. The symbol $x$ can be equivalently represented by the $2n$-bit binary tuple

$$(\underline{x}_R, \underline{x}_I) = (x_R(1), x_R(2), \ldots, x_R(n), x_I(1), x_I(2), \ldots, x_I(n)). \tag{7}$$

Note that the above channel inputs satisfy a unit energy constraint at each discrete-time, and are therefore valid inputs even for the Gaussian network with a unit power constraint. This property will be helpful in proving the approximate optimality of the linear network code.

The channel gains are unchanged from the Gaussian model. As in the Gaussian model, the channel between two nodes $i$ and $j$ in the discrete network simply multiplies the input $x_i$ by the corresponding channel gain $h_{ij}$. At a receiver, the received signal is defined through the composition of the following operations:

- First add all the incoming signals by the standard summation over $\mathbb{C}$.
- Then discard the signs of the real and imaginary parts.
- Further discard the fractional portions of the real and imaginary parts and retain only the integer portion.
- Then quantize the integer portion of the real and imaginary parts by truncating their binary expansions to $n$ bits of precision, where $n$ is as defined in (3).

Thus $y = y_R + \imath y_I$ is the received signal at a node in the Gaussian model, and we denote the binary expansions of the integer parts of $|y_R|$ and $|y_I|$ by $\sum_{k=1}^{\infty} 2^k y_R(k)$ and $\sum_{k=1}^{\infty} 2^k y_I(k)$, respectively. The received signal in the discrete network is then

$$y^{'} := [y] := \left(\sum_{k=1}^{n} 2^k y_R(k)\right) + \imath \left(\sum_{k=1}^{n} 2^k y_I(k)\right). \tag{8}$$

As with the transmit signals, $y^{'}$ can be equivalently described by the $2n$-bit binary tuple

$$(\underline{y}_R^{'}, \underline{y}_I^{'}) \;=\; (y_R(1), y_R(2), \ldots, y_R(n), y_I(1), y_I(2), \ldots, y_I(n)). \tag{9}$$

We will use the compact notation $[\cdot]$ to represent the overall quantization operation:

$$y_j^{'} \;:=\; [y_j] \;:=\; [\sum_{i\in\mathcal{N}(j)} h_{ij}x_i + z_j]. \tag{10}$$

It is important to note that each received signal in the discrete network can be obtained from the corresponding received signal in the Gaussian network by performing elementary quantization operations (when their transmitted signals are identical, as we intend to be the case). In fact, since the transmit signals in the discrete network are valid transmit signals for the Gaussian network, we use the same notation for the transmit signals in both models.

In a similar way, one also obtains a discrete MIMO network corresponding to a MIMO Gaussian relay network. In the MIMO discrete network, each transmit and receive antenna can be treated as a virtual node. As before, every transmit and received signal (corresponding to every transmit or receive antenna) is quantized to lie in a finite set, and the granularity of the quantization will take into account all the channel gains between various antennas in the network. The transmit signals lie in a finite set and can be described by a $2n_{MIMO}$-bit tuple. The receive signals are quantized and described by a $2n_{MIMO}$-bit tuple, where

$$n_{MIMO} \;:=\; \max_{(i,j)\in\mathcal{E}} \max_{\substack{k=1,\ldots,T_i,\\ l=1,\ldots,U_j}} \{\lfloor \log|h_{ijR}^{kl}|\rfloor, \lfloor \log|h_{ijI}^{kl}|\rfloor\}. \tag{11}$$

## III. The cut-set bound on the capacity of relay networks

The cut-set bound [7], [20] on the capacity $C$ of a relay network is

$$C \;\leq\; \max_{p(x_0,x_1,\ldots,x_{M-1})} \min_{\Omega\in\Lambda} I(x_\Omega; y_{\Omega^c}|x_{\Omega^c}), \tag{12}$$

where $\Lambda$ is the set of partitions of $\{0,1,\ldots,M\}$ with the source $0 \in \Omega$ and the destination $M \in \Omega^c$, and $\{x_i\}$ and $\{y_j\}$ denote the transmit and receive signals. Additionally, in the Gaussian model, the inputs must satisfy an average power constraint, $E[|x_i|^2] \;\leq\; 1, \forall i$. Let $CS_G$ be the cut-set bound of the Gaussian network and let $CS_D$ be the cut-set bound of the corresponding discrete network. In this section, we start the program of approximating Gaussian networks with the discrete model by proving that the cut-set bounds of the two networks are within a bounded gap. The importance of this result will become clear in the later sections when we prove that the cut-set bound is achievable for the layered discrete network up to a constant gap with a simple linear coding scheme, and the same scheme can be used on the layered Gaussian network by merely quantizing the signals.

*Lemma 3.1: The cut-set bounds of the Gaussian network and the corresponding discrete network are within a constant gap of $O(M \log M)$ bits,*

$$|CS_G - CS_D| \;=\; O(M \log M), \tag{13}$$

*with the gap independent of channel gains or SNR.*

*Proof:* The lemma is proved in two steps: $CS_G \;\geq\; CS_D$ in Section III-1 and $CS_G - CS_D \;=\; O(M \log M)$ in Section III-2. The procedure to prove both inequalities are the same. We consider a particular cut $\Omega$ in the network and choose an input distribution for this cut in one of the models. Then, in a series of steps, we transform the channel inputs and outputs of the cut to the corresponding channel inputs and outputs in the other model. We then bound the loss in the mutual information in this transformation. Repeating this procedure across all the cuts in the network completes the proof.

*1) $CS_G \;\geq\; CS_D$:* The transmit signals in the discrete network are a strict subset of the valid inputs for the Gaussian network, and the received signals in the discrete network are obtained by quantizing the corresponding received signals in the Gaussian network. Hence, as noted above, any operation in the discrete network can be simulated on the Gaussian network.

Choose any input distribution for the transmit signals in the discrete relay network. Retain the same distribution for the inputs in the Gaussian network. Since the received signals in the discrete network are obtained by quantizing the received signals in the Gaussian network, for any cut $\Omega$, by the data processing lemma [20],

$$I(x_\Omega; y_{\Omega^c} | x_{\Omega^c}) \;\geq\; I(x_\Omega; y'_{\Omega^c} | x_{\Omega^c}). \tag{14}$$

This proves that $CS_G \geq CS_D$.

*2) $CS_G - CS_D \;=\; O(M \log M)$:* We show that the mutual information across a cut in the discrete network is at least as high as the mutual information across the same cut in the Gaussian network, up to $O(M \log M)$ bits.

*Step 1: Cut-by-cut analysis*

The cut-set bound for the Gaussian network is

$$\begin{aligned} CS_G &= \max_{p(x_0,x_1,\ldots,x_{M-1})} \min_{\Omega\in\Lambda} I(x_\Omega; y_{\Omega^c}|x_{\Omega^c}) && (15)\\ &\leq \min_{\Omega\in\Lambda} \max_{p(x_0,x_1,\ldots,x_{M-1})} I(x_\Omega; y_{\Omega^c}|x_{\Omega^c}). && (16)\end{aligned}$$

We consider a particular cut $\Omega/\Omega^c$ in the network. The received signal at the $j$-th node is given by

$$y_j = \sum_{i\in N(j)} h_{ij}x_i + z_j. \tag{17}$$

The mutual information across the cut, $\mathcal{I}_1 := I(x_\Omega; y_{\Omega^c}|x_{\Omega^c})$, is maximized by the choice of jointly Gaussian inputs.

*Step 2: Positive fractional inputs*

Instead of the optimal joint Gaussian distribution for the inputs that maximizes $\mathcal{I}_1$, we choose a different input distribution. Consider an input symbol $x = \frac{1}{\sqrt{2}}(x_R + \imath\, x_I)$, where $x_R$ and $x_I$ are independent and uniformly distributed on $(0,1)$, i.e., $x_R$ and $x_I$ are positive fractions. Since $E[|x|^2] = 1/3$, it satisfies the average power constraint for the Gaussian network's channel inputs. Each input in the network is chosen independently and with the same distribution as $x$, and is denoted by $\{x_i^{(2)}\}$. The received signals are denoted by $\{y_j^{(2)}\}$. In Lemma A.1 in Section A, it is shown that the loss in the mutual information for this choice of inputs is $O(M)$. The mutual information of this channel is $\mathcal{I}_2 := I(x_\Omega^{(2)}; y_{\Omega^c}^{(2)}|x_{\Omega^c}^{(2)})$ and compares to the channel in Step 1 as

$$\mathcal{I}_1 - \mathcal{I}_2 = O(M). \tag{18}$$

*Step 3: Quantization of the received signal*

Next we quantize the received signal as follows:

- Retain only the integer portions of the real and imaginary parts, and discard the fractional portions, which we denote by $d_j^{(3)}$. The real and imaginary parts of $d_j^{(3)}$ have the same sign as the signal.
- Discard the signs of the real and imaginary parts, which we denote by $s_{jR}^{(3)}$ and $s_{jI}^{(3)}$, respectively.

We denote the quantized received signal by $y_j^{(3)}$. The mutual information across the channel in Step 2 can be rewritten as

$$\begin{aligned}\mathcal{I}_2 &= I(x_\Omega^{(2)}; y_{\Omega^c}^{(3)}, d_{\Omega^c}^{(3)}, s_{\Omega^c}^{(3)} | x_{\Omega^c}^{(2)}) && (19)\\ &\leq I(x_\Omega^{(2)}; y_{\Omega^c}^{(3)} | x_{\Omega^c}^{(2)}) + I(x_\Omega^{(2)}; d_{\Omega^c}^{(3)} | x_{\Omega^c}^{(2)}, y_{\Omega^c}^{(3)}) + H(s_{\Omega^c}^{(3)}) && (20)\\ &\leq I(x_\Omega^{(3)}; y_{\Omega^c}^{(3)} | x_{\Omega^c}^{(2)}) + I(x_\Omega^{(2)}; d_{\Omega^c}^{(3)} | x_{\Omega^c}^{(2)}, y_{\Omega^c}^{(3)}) + 2|\Omega^c|. && (21)\end{aligned}$$

Now the fractional part $d_j^{(3)}$ is given by adding the fractional part of the signal with the fractional part of the noise.

$$\begin{aligned}I(x_\Omega^{(2)}; d_{\Omega^c}^{(3)} | x_{\Omega^c}^{(2)}, y_{\Omega^c}^{(3)}) &= h(d_{\Omega^c}^{(3)} | x_{\Omega^c}^{(2)}, y_{\Omega^c}^{(3)}) - h(d_{\Omega^c}^{(3)} | x_{\Omega^c}^{(2)}, y_{\Omega^c}^{(3)}, x_\Omega^{(2)}) && (22)\\ &\leq \sum_{j \in \Omega^c} h(d_j^{(3)}) - h(\tilde{z}_j^{(3)}), && (23)\end{aligned}$$

where $\tilde{z}_j^{(3)}$ is the fractional part of the Gaussian noise. Since $d_j^{(3)}$ is fractional with $E[|d_j^{(3)}|^2] \leq 2$, its differential entropy is upper bounded by that of a Gaussian distribution with variance $2$, and so $h(d_j^{(3)}) \leq \log(2\pi e)$. Since $z_j$ is distributed as $\mathcal{CN}(0,1)$, $h(\tilde{z}_j^{(3)}) = O(1)$. Hence,

$$I(x_\Omega^{(2)}; d_{\Omega^c}^{(3)} | x_{\Omega^c}^{(2)}, y_{\Omega^c}^{(3)}) \leq |\Omega^c|(\log(2\pi e) - O(1)) = O(M). \quad (24)$$

So, defining $\mathcal{I}_3 := I(x_\Omega^{(2)}; y_{\Omega^c}^{(3)} | x_{\Omega^c}^{(2)})$, we have

$$\mathcal{I}_2 - \mathcal{I}_3 = O(M). \quad (25)$$

*Step 4: Further quantization of the received signal*

Next we further quantize the received signal at the end of the previous step by restricting the binary expansions of its real and imaginary parts to $n$ bits, and denote the result by $y_j^{(4)}$, where

$$n := \max_{(i,j) \in \mathcal{E}} \max\{\lfloor \log |h_{ijR}| \rfloor, \lfloor \log |h_{ijI}| \rfloor\}. \quad (26)$$

Note that we have previously quantized the real and imaginary parts to be positive integers. Denote the discarded part of the received signal by $d_j^{(4)}$. The mutual information of the channel in Step 3 can be rewritten as

$$\begin{aligned}\mathcal{I}_3 &= I(x_\Omega^{(2)}; y_{\Omega^c}^{(4)}, d_{\Omega^c}^{(4)} | x_{\Omega^c}^{(2)}) && (27)\\ &= I(x_\Omega^{(2)}; y_{\Omega^c}^{(4)} | x_{\Omega^c}^{(2)}) + I(x_\Omega^{(2)}; d_{\Omega^c}^{(4)} | x_{\Omega^c}^{(2)}, y_{\Omega^c}^{(4)}) && (28)\\ &\leq I(x_\Omega^{(2)}; y_{\Omega^c}^{(4)} | x_{\Omega^c}^{(2)}) + I(x_\Omega^{(2)}; d_{\Omega^c}^{(4)} | x_{\Omega^c}^{(2)}). && (29)\end{aligned}$$

We bound $I(x_\Omega^{(2)}; d_{\Omega^c}^{(4)} | x_{\Omega^c}^{(2)})$ next. From the definition of $n$, $|h_{ij}| = \sqrt{(h_{ijR}^2 + h_{ijI}^2)} \leq 2^{n+1}$. Since $|x_i^{(2)}| \leq 1$, $|\sum_{i \in N(j)} h_{ij} x_i^{(2)}| \leq M 2^{n+1}$. Hence, the binary expansion of the integer part of $|\sum_{i \in N(j)} h_{ij} x_i^{(2)}|$ has $(n + O(\log M))$ bits. Since $d_j^{(4)}$ is the portion of the received signal that exceeds $n$ bits of representation, it is easy to see that at most $O(\log M)$ higher order bits in the binary representation of $\sum_{i \in N(j)} h_{ij} x_i^{(2)}$ influence $d_j^{(4)}$. Therefore, we have $I(x_\Omega^{(2)}; d_j^{(4)} | x_{\Omega^c}^{(2)}) = O(\log M)$ and subsequently $I(x_\Omega^{(2)}; d_{\Omega^c}^{(4)} | x_{\Omega^c}^{(2)}) = O(M \log M)$. Now define $\mathcal{I}_4 := I(x_\Omega^{(2)}; y_{\Omega^c}^{(4)} | x_{\Omega^c}^{(2)})$ and we get

$$\mathcal{I}_3 - \mathcal{I}_4 = O(M \log M). \tag{30}$$

*Step 5: Quantization of the transmit signals*

Next, we restrict the real and imaginary parts of the scaled inputs also to $n$ bits. Let the binary expansion of $\sqrt{2}\, x_{iR}^{(4)}$ be $0.x_{iR}(1)x_{iR}(2)\ldots$, so

$$x_{iR}^{(4)} = \frac{1}{\sqrt{2}} \sum_{p=1}^{\infty} 2^{-p} x_{iR}(p). \tag{31}$$

Similarly denote the binary expansion of $\sqrt{2}\, x_{iI}^{(4)}$. Define

$$x_{iR}^{'} := \frac{1}{\sqrt{2}} \sum_{p=1}^{n} x_{iR}(p) 2^{-p},$$

$$x_{iI}^{'} := \frac{1}{\sqrt{2}} \sum_{p=1}^{n} x_{iI}(p) 2^{-p}.$$

We will consider the new inputs $x_i^{'} := x_{iR}^{'} + \imath x_{iI}^{'}$ and let the corresponding received signals under these inputs be denoted by $y_j^{'}$. The mutual information here compares against that in Step $4$ as

$$\mathcal{I}_4 = I(x_\Omega^{(2)}, x_\Omega^{'}; y_{\Omega^c}^{(4)} | x_{\Omega^c}^{(2)}, x_{\Omega^c}^{'}), \quad \text{(since } x_i^{'} \text{ is only a function of } x_i^{(2)}\text{)} \tag{32}$$

$$\leq I(x_\Omega^{(2)}, x_\Omega^{'}; y_{\Omega^c}^{(4)}, y_{\Omega^c}^{'} | x_{\Omega^c}^{(2)}, x_{\Omega^c}^{'}) \tag{33}$$

$$= I(x_\Omega^{(2)}, x_\Omega^{'}; y_{\Omega^c}^{'} | x_{\Omega^c}^{(2)}, x_{\Omega^c}^{'}) + I(x_\Omega^{(2)}, x_\Omega^{'}; y_{\Omega^c}^{(4)} | x_{\Omega^c}^{(2)}, x_{\Omega^c}^{'}, y_{\Omega^c}^{'}) \tag{34}$$

$$= I(x_\Omega^{'}; y_{\Omega^c}^{'} | x_{\Omega^c}^{'}) + I(x_\Omega^{(2)}, x_\Omega^{'}; y_{\Omega^c}^{(4)} | x_{\Omega^c}^{(4)}, x_{\Omega^c}^{'}, y_{\Omega^c}^{'}) \tag{35}$$

$$\leq I(x_\Omega^{'}; y_{\Omega^c}^{'} | x_{\Omega^c}^{'}) + H(y_{\Omega^c}^{(4)} | y_{\Omega^c}^{'}). \tag{36}$$

In (35), the first mutual information term is obtained due the Markov chain $\{x_i^{(2)}\} \rightarrow \{x_i^{'}\} \rightarrow \{y_j^{'}\}$. Since

$$|h_{ij}(x_i^{(2)} - x_i^{'})| \;=\; |h_{ij}||x_i^{(2)} - x_i^{'}| \tag{37}$$

$$\leq\; (2^{n+1})(2^{-n}) \;\leq\; 2, \tag{38}$$

we get

$$|\sum_{i \in N(j)} h_{ij}(x_i^{(2)} - x_i^{'})| \;=\; O(M). \tag{39}$$

Hence the binary representation of the integer part of $|\sum_{i\in N(j)} h_{ij}(x_i^{(2)} - x_i^{'})|$ will have $O(\log M)$ bits. This results in $H(|y_j^{(4)} - y_j^{'}|) \;=\; O(\log M)$, due to which $H(y_{\Omega^c}^{(4)}|y_{\Omega^c}^{'}) \;=\; O(M \log M)$ bits. We obtain

$$\mathcal{I}_4 - I(x_{\Omega}^{'}; y_{\Omega^c}^{'}|x_{\Omega^c}^{'}) \;=\; O(M \log M). \tag{40}$$

Now observe that $x_i^{'}$ and $y_j^{'}$ are the transmit and receive signals in the discrete network. Also, $\{x_i^{'}\}$ are i.i.d. and uniformly distributed on their alphabet. In Step 1, we started with the optimal jointly Gaussian inputs for the cut in the Gaussian network and concluded in this step with i.i.d. uniform inputs for the discrete network. The total loss in the mutual information in this transformation is $O(M \log M)$. Now define $\overline{CS}_D$ as

$$\overline{CS}_D \;:=\; \min_{\Omega \in \Lambda} \; I(x_{\Omega}^{'}; y_{\Omega^c}^{'}|x_{\Omega^c}^{'}). \tag{41}$$

$\overline{CS}_D$ evaluates the right hand side of the cut-set bound (12) for the discrete network for a specific choice of the input distribution. As a consequence of the above arguments, we have also proved that $CS_G - \overline{CS}_D \;=\; O(M \log M)$. ■

The choice of i.i.d. uniform inputs for the discrete network will be useful in Section IV-C.

## IV. Linear network code for discrete relay networks

In the Section II-B we have presented a quantizaton-based digital interface for operating a Gaussian relay network, which we have called the *discrete network.* It has the properties that it is obtained by merely quantizing the signals at each node in the Gaussian relay network, and that its cut-set bound on the capacity is nearly that of the original Gaussian relay network. Therefore any coding strategy for the discrete network that achieves rates close to the cut-set bound can be easily converted to a nearly capacity-acheiving strategy for the Gaussian relay network by simply quantizing the signals. This motivates and sets the stage for the problem of determining nearly capacity-achieving coding strategies for the discrete network, which is the topic of this section. What we determine is that simple linear network coding on the bits of the en bloc finite representations of the received signals will suffice for the discrete network. The linear network code presented in the sequel achieves all rates within $O(M^2)$ bits of the cut-set bound of the layered discrete relay network, with the constant gap independent of channel gains and SNR, and only a function of the number of nodes in the network.

### A. *Coding scheme for relay networks*

Any coding scheme for a relay network requires specification of the source's codewords, the functions applied each time to the received signals by the relay nodes, and the destination's decoding function.

A $(2^{NR}, N)$ code for a relay network is an encoding function for the source

$$\underline{x}_0 : \ \{1, 2, \ldots, 2^{NR}\} \to \mathcal{X}^N,$$

where $\mathcal{X}$ is the input alphabet of the channel, and a set of encoding functions for relay node $k$,

$$g_{k,m} : \mathcal{Y}^{m-1} \to \mathcal{X}, \ \text{for } m = 1, 2, \ldots, N, \ k = 1, 2, \ldots, M-1,$$

where $\mathcal{Y}$ is the alphabet of node $k$'s received signal. For simplicity of exposition, we assume that the input alphabets of all the relay nodes are the same, and that so are their output alphabets. The symbols transmitted by a relay can depend on all the symbols previously received by the relay. The decoding function of the destination $M$ is given by

$$g_M : \mathcal{Y}^N \to \{1, 2, \ldots, 2^{NR}\}.$$

Let $\mathcal{W}$ be a random variable uniformly distributed on $\{1, 2, \ldots, 2^{NR}\}$ that corresponds to the message that source $0$ wants to communicate. Such a $\mathcal{W}$ is mapped to the codeword $\underline{x}_0(\mathcal{W})$. The average probability of error is given by

$$P_e \ = \ Pr(g_M(\underline{y}_M) \ \neq \ \mathcal{W}).$$

The capacity of the relay network is the supremum of all rates $R$ such that for any $\epsilon > 0$, there exists a block length $N$ and a coding strategy for which $P_e < \epsilon$.

Since the discrete network can be simulated on the Gaussian network, any coding scheme for the discrete relay network can be directly used on the Gaussian relay network. We formalize this notion in the following lemma.

*Lemma 4.1: A coding scheme for the discrete relay network can be lifted to the original Gaussian relay network with no change to its structure and no loss in the rate.*

*Proof:* The proof is based on the observation that the discrete network can be simulated on the Gaussian network. As mentioned in Sec. II-B, the discrete network is obtained by constraining the transmit signals in the Gaussian network to a finite set and quantizing the received signals in the Gaussian network to a finite subset. Hence, any coding scheme for the discrete network can be directly employed on the Gaussian network by operating the Gaussian network on the quantization-based interface defined by the discrete network. This procedure does not require any changes to the coding scheme and does not decrease the rate of the coding scheme. This result can be extended to codes for MIMO relay networks and multicast networks.

We have not explicitly defined a coding scheme for a general Gaussian network, but the above ideas are applicable to a general Gaussian network too. ■

### B. Layered networks

In the coding scheme described above, the encoding functions $\{g_{k,m}\}$ at relay nodes operate on a symbol-by-symbol basis and can vary with time. We restrict our attention to layered discrete networks, for which the encoding functions $\{g_{k,m}\}$ at the relay nodes can be simplified. In a layered network [2], as the name suggests, nodes are divided into groups or layers. The nodes of one layer can only transmit to nodes of the subsequent layer. The source node $0$ is the sole node in the zeroeth layer, and the last layer, layer $L$, contains only the destination $M$. All other relays are divided among the intermediate layers. The nodes in layer $k$ are denoted by $\mathcal{L}_k$. An example with four layers of nodes is shown in Figure 1 with $\mathcal{L}_0 = \{0\}, \mathcal{L}_1 = \{1, 2\}, \mathcal{L}_2 = \{3, 4\}$, $\mathcal{L}_3 = \{5\}$, and $L = 3$.

In a layered network we can restrict attention to the following simplified block-by-block coding scheme, where each block consists of $N$ symbols. Consider a $(2^{NR}, N)$ code for the relay network, and suppose that the source transmits a codeword of length $N$, which we call a "block." Now let all the relays in $\mathcal{L}_1$ buffer their received signals for $N$ time units. Subsequently each relay generates $N$-length transmit vectors, i.e., blocks, as a function of their received vectors.

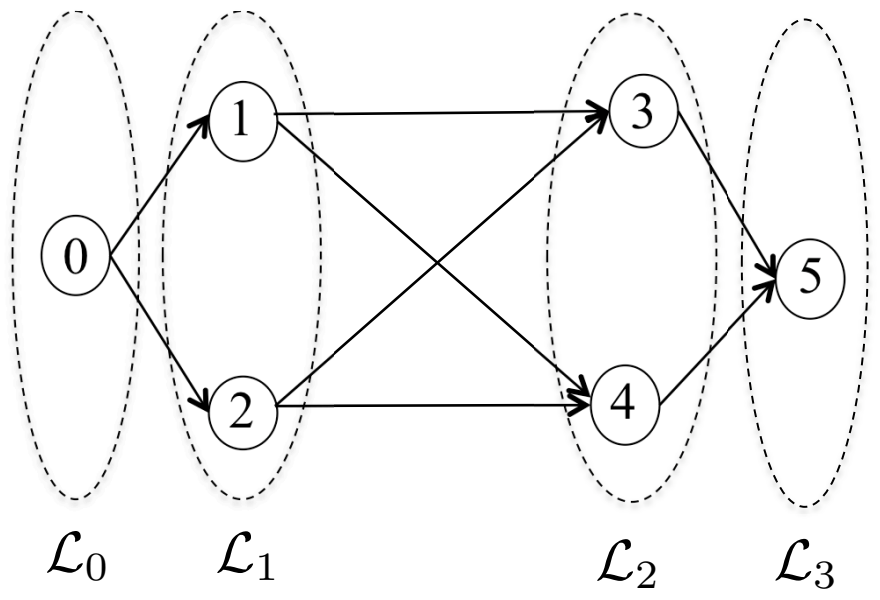

Fig. 1. Example of a layered network.

This is possible since only the source transmits to the relays in $\mathcal{L}_1$, and these relays generate their transmit symbols causally as a function of their previous receptions. Similarly, each node in $\mathcal{L}_k$ waits for the nodes in $\mathcal{L}_{k-1}$ to complete their transmissions, buffers the $N$ received signals, and then transmits a block of length $N$. Finally, the destination receives the block of $N$ symbols and attempts to decode the source's transmission. There is no loss in the capacity in operating on a block-by-block basis. The source can start the transmission of the next codeword once it completes the transmission of the first codeword. Also the relays in $\mathcal{L}_1$ are full-duplex, they transmit and receive at the same time. So the relays can continue receiving the next block of symbols while they transmit a block of symbols. The destination will continue to receive blocks of symbols one after another. In this way, we achieve a constant rate $R$ of communication in the network.

### *C. Linear network codes for the discrete relay networks*

The linear network code we propose is constructed on the discrete network. As described in Section II-B, the transmit and receive signals in the discrete network are quantized to lie in a finite set of complex numbers consisting of real and complex parts in [0,1], each with only $n$ bits of precision.

Concerning transmissions, the transmit signal $x_i = x_{i,R} + \imath x_{i,I}$ has real and imaginary parts

given by

$$x_{i,R} = \sum_{k=1}^{n} 2^{-k} x_{i,R}(k), \tag{42}$$

$$x_{i,I} = \sum_{k=1}^{n} 2^{-k} x_{i,I}(k). \tag{43}$$

with each $x_{i,R}(i)$ and $x_{i,I}(j)$ in $\mathbb{F}_2$, and where

$$n := \max_{(i,j) \in \mathcal{E}} \max\{\lfloor \log |h_{ijR}| \rfloor, \lfloor \log |h_{ijI}| \rfloor\}. \tag{44}$$

The received signals in the discrete network $\{y_j^{'}\}$ are obtained by quantizing the received signals in the Gaussian network. The Gaussian received signal is quantized by discarding the signs of the real and imaginary parts, further discarding the fractional portions of the real and imaginary parts and retaining only the integer portion, and finally quantizing the integer portion of the real and imaginary parts by truncating their binary expansions to $n$ bits of precision. If $y_j = y_{j,R} + \imath y_{j,I}$ is the Gaussian received signal, and we denote the binary expansions of the integer parts of $|y_{j,R}|$ and $|y_{j,I}|$ by $\sum_{k=1}^{\infty} 2^k y_{j,R}(k)$ and $\sum_{k=1}^{\infty} 2^k y_{j,I}(k)$, respectively, then the received signal $y_j^{'}$ in the discrete network is

$$y_j^{'} = [y_j] := \left( \sum_{k=1}^{n} 2^k y_{j,R}(k) \right) + \imath \left( \sum_{k=1}^{n} 2^k y_{j,I}(k) \right). \tag{45}$$

Hence we regard the received signal in the discrete network as a binary tuple

$$(y_{j,R}(1), y_{j,R}(2), \ldots, y_{j,R}(n), y_{j,I}(1), y_{j,I}(2), \ldots, y_{j,I}(n))$$

of length $2n$ where each entry is 0 or 1. Similarly, the transmitted signal in the discrete network can also be so regarded. In the rest of this paper, we reserve the phrase "binary $2n$-tuple" or "$2n$-tuple" to describe such a vector, which can be converted in a straightforward fashion to a complex symbol for transmission, or can be obtained in a straightforward fashion from a complex symbol that is received.

After $N$ such symbols have been received, there is a $2nN$-length binary vector that is the received block, and similarly there is a $2nN$-length binary vector that is the transmitted block. We will reserve the phrase "binary $2nN$-vector" or "$2nN$-vector" to refer to such a vector which represents a block of $N$ received symbols, either for block transmission or after block reception. We will represent a $2nN$-length received or transmitted binary vector at a node $j$ using an underbar, as in $\underline{y}_j^{'}$ or $\underline{x}_j$, respectively.

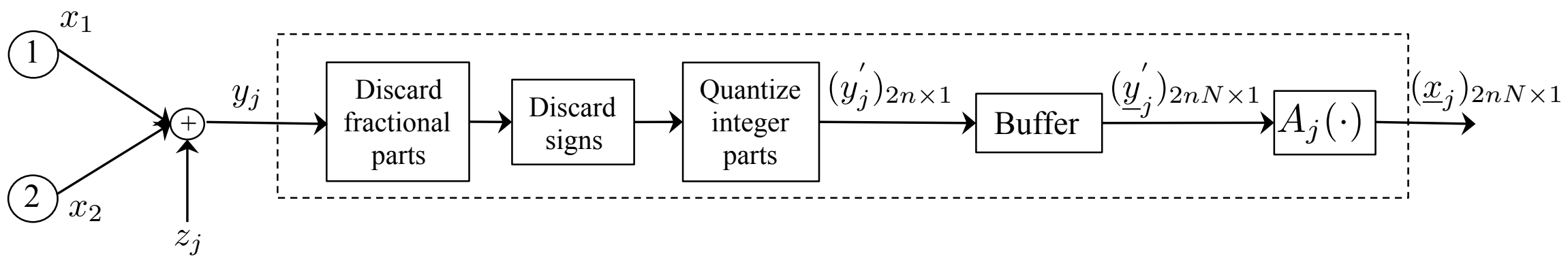

Fig. 2. Linear encoding at node $j$.

We will employ a linear coding scheme where the transmitted block is simply obtained by multiplying the received block by a $2nN \times 2nN$ matrix of 0s and 1s.

The overall coding scheme is randomly generated, and is simple to describe.

*1)* **Source's codewords***:* There are $2^{NR}$ messages. Using a uniform distribution on binary $2n$-tuples, the source randomly generates a set of $2^{NR}$ codewords, each of length $2nN$, where each codeword is constructed by independently picking $N$ binary $2n$-tuples from the uniform distribution. Note that each codeword corresponds to $N$ complex symbols, which are transmitted over $N$ discrete time instants. The choices of the rate $R$ and the block-length $N$ are elaborated later in the proof of Lemma 4.2 in Section IV-D. The source transmits the codeword corresponding to the particular message that has been chosen for communication.

*2)* **Relay's linear mappings***:* Relay $j$ randomly chooses a $2nN \times 2nN$ binary matrix $A_j$, by independently picking each entry as either $0$ or $1$ with equal probability. This will be the matrix representing the linear code at node $j$. The relay buffers $N$ received binary $2n$-tuples, and adjoins them to construct a binary $2nN$-vector that constitutes the received block $\underline{y}_j^{'}$. It then multiplies this binary $2nN$-vector by $A_j$ to obtain a binary $2nN$-vector $\underline{x}_j$ that constitutes the transmit block. It splits this binary $2nN$-vector into $N$ binary $2n$-tuples. Converting each of the binary $2n$-tuples back into complex numbers, with the real and imaginary parts, each in $[0, 1]$, and each of $n$-bit precision, gives $N$ complex symbols. These $N$ complex transmit symbols are transmitted by the relay node over $N$ discrete-time instants. An example of the encoding operation at node $j$ is shown in Figure 2.

All the $M - 1$ relays independently construct their binary encoding matrices in a similar manner as described above.

Next, to define decoding, we define strong typicality of vectors:

*Definition 1:* A vector $\underline{x} \in \mathcal{X}^N$ is defined to be $\epsilon$-*strongly typical* with respect to a distribution

$p(x)$, denoted by $\underline{x} \in \mathcal{T}_{\epsilon,p}$, if

$$|\nu_x(\underline{x}) - p(x)| \ \leq \ \epsilon\, p(x), \ \forall\, x \in \mathcal{X}, \tag{46}$$

where $\epsilon \in \mathcal{R}^+$ and $\nu_x(\underline{x}) = \frac{1}{N}|n : x_n = x|$ is the empirical frequency.

This is extended in the standard way to include joint strong typicality of vectors (see [20]). Vectors $\underline{x} = (x_1, x_2, \ldots, x_N) \in \mathcal{X}^N$ and $\underline{y} = (y_1, y_2, \ldots, y_N) \in \mathcal{Y}^N$ are $\epsilon$-jointly strongly typical with respect to the distribution $p(x, y)$, if $\underline{z} = (z_1, z_2, \ldots, z_N)$, where each $z_i = \begin{pmatrix} x_i \\ y_i \end{pmatrix}$ is jointly typical with respect to the distribution $p(x, y)$. We employ the following definitions from [23], slightly modifying them to include the linear operations.

Define a singleton set $\chi_0(w) := \{\underline{x}_0(w)\}$ for every message $w$.

Now consider a node $j$ in $\mathcal{L}_1$, the first layer. Define the *set of received vectors at node $j$ associated with a message $w$* as

$$\mathcal{Y}_j^{'}(w) \ := \ \{\underline{y}_j^{'} : (\underline{y}_j^{'}, \underline{x}_0(w)) \in \mathcal{T}_{\epsilon,p}\}, \tag{47}$$

with $p$ denoting the distribution $p(\tilde{x}_0, \tilde{y}_j^{'})$ where $\tilde{x}_0$ is uniformly distributed and $p(\tilde{y}_j^{'}|\tilde{x}_0)$ models the channel from node $0$ to node $j$ in the discrete network.

Continuing to consider the node $j \in \mathcal{L}_1$, we define the *set of transmit vectors at node $j$ that are associated with the message $w$* as

$$\chi_j(w) = \{\underline{x}_j : \underline{x}_j \ = \ A_j(\underline{y}_j^{'}), \text{ where } \underline{y}_j^{'} \in \mathcal{Y}_j^{'}(w)\}. \tag{48}$$

Next consider a node $j \in \mathcal{L}_k$ with $k \geq 2$. Noting that any $i \in \mathcal{N}(j)$ belongs to layer $\mathcal{L}_{k-1}$, we recursively (in $k$) define its set of received vectors at node $j$ associated with message $w$ as

$$\mathcal{Y}_j^{'}(w) \ := \ \{\underline{y}_j^{'} : (\underline{y}_j^{'}, \{\underline{x}_i\}_{i \in \mathcal{N}(j)}) \in \mathcal{T}_{\epsilon,p}, \text{ for some } \underline{x}_i \in \chi_i(w), \text{ for each } i \in \mathcal{N}(j)\}. \tag{49}$$

The distribution $p$ above is $p(\{\tilde{x}_i\}_{i \in \mathcal{N}_j}, \tilde{y}_j^{'})$, where $\{\tilde{x}_i\}_{i \in \mathcal{N}(j)}$ are independent and uniformly distributed, and $p(\tilde{y}_j^{'}|\{\tilde{x}_i\}_{i \in \mathcal{N}(j)})$ models the channel from the nodes in $\mathcal{N}(j)$ to $j$ in the discrete network. We also define the set of transmitted vectors at node $j$ associated with the message $w$ as

$$\chi_j(w) := \{\underline{x}_j : \underline{x}_j \ = \ A_j(\underline{y}_j^{'}), \text{ for some } \underline{y}_j^{'} \in \mathcal{Y}_j^{'}(w)\}. \tag{50}$$

*Definition 2:* We write $(\underline{y}_j^{'}, w) \in \mathcal{T}^{\epsilon}$ if $\underline{y}_j^{'} \in \mathcal{Y}_j^{'}(w)$.

*3)* **Decoding at the destination***:* The destination receives the $2nN$-vector $\underline{y}_M^{'}$, and decodes by searching for a message $w$ such that $(\underline{y}_M^{'}, w) \in \mathcal{T}^{\epsilon}$. If it finds more than one such message or if it finds none, it declares an error. Else it declares the unique choice as its estimate of the source's message.

It should be noted that though the relay's encoding is a linear operation over the block of $2nN$-vectors, the end-to-end channel from the binary $2nN$-vectors transmitted by the source to the block of binary $2nN$-vectors received by the destination is not linear over the binary field, and we cannot describe decoding as just inverting a binary matrix $2nN \times 2nN$ matrix. This is because of the truncation operations that are an integral part of the very definition of the discrete network.

### *D. Computing the rate achieved by the linear network code*

Next we compute the rate achievable by the linear network code. In (41), we defined $\overline{CS}_D$ as the cut-set bound's value for the discrete network for a specific choice of the input distribution.

*Lemma 4.2: The linear network code achieves all rates within a gap of $O(M^2)$ bits of $\overline{CS}_D$ for layered networks, where the $O(M^2)$ gap is independent of channel gains or SNR.*

*Proof:*

Probability of error: The probability of error for the coding scheme (see Section IV-A for the notation) is given by

$$P_e = \Pr(g_M(\underline{y}_M^{'}) \neq \mathcal{W}) \tag{51}$$

$$= \frac{1}{2^{NR}} \sum_{w=1}^{2^{NR}} \Pr(g_M(\underline{y}_M^{'}) \neq w | \mathcal{W} = w). \tag{52}$$

As is standard, since $P_e$ is symmetric in the transmitted message, we assume wlog that $\mathcal{W} = 1$ and evaluate the probability of error when the first codeword is transmitted, $P_e|_{\mathcal{W}=1}$.

Error events: The possible error events at the destination $M$ are

- $E_0$: One of the transmitted or received vectors in the network is not strongly typical. That is $\underline{y}_j^{'} \notin \mathcal{T}_{\epsilon,p}$, where $p$ denotes the distribution induced on $y_j^{'}$ by the uniform distribution on each $x_i$ for $i \in \mathcal{N}(j)$, and by the conditional probability $p(y_j^{'}|\{x_i\}_{i\in\mathcal{N}(j)})$ describing the channel in the discrete network from the nodes in $\mathcal{N}(j)$ to $y_j^{'}$.

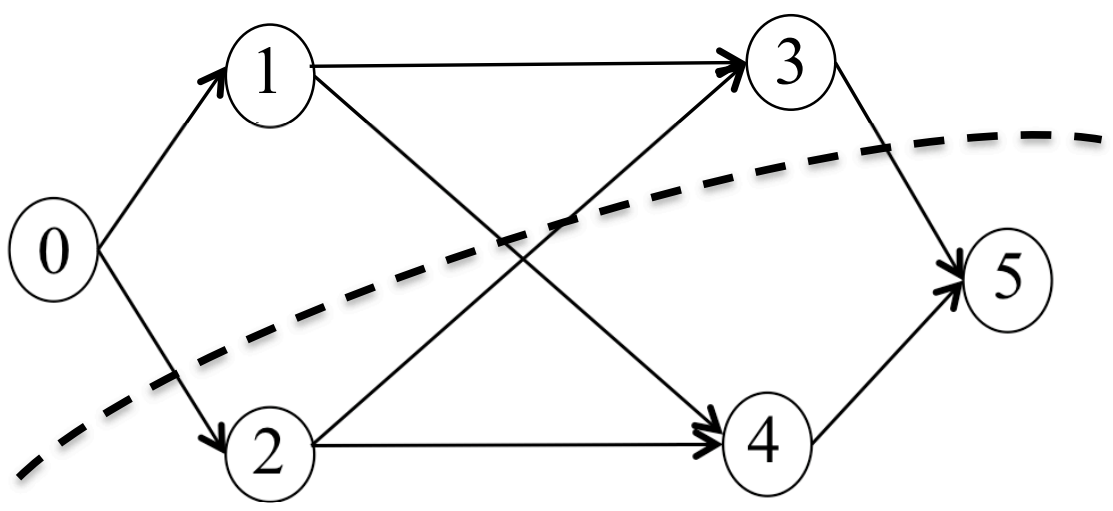

Fig. 3. Example of a cut $\Omega$ in the network in Figure 1. Here $\Omega = \{0, 1, 3\}$ and $\Omega^c = \{2, 4, 5\}$.

- $E_1$: $(\underline{y}^{'}_{M}, 1) \notin \mathcal{T}^\epsilon$.
- $E_w$: $(\underline{y}^{'}_{M}, w) \in \mathcal{T}^\epsilon$, where $w \neq 1$.

By applying the union bound,

$$P_e|_{\mathcal{W}=1} \;\; \leq \;\; \Pr(E_0) + \Pr(E_1 \wedge E_0^c) + \Pr(\bigvee_{w \neq 1} E_w \wedge E_0^c \wedge E_1^c). \tag{53}$$

From Lemma B.1 (see Section B), for any $\epsilon > 0$,

$$\Pr(E_0) + \Pr(E_1) \;\; \leq \;\; \epsilon, \text{ for } N \text{ sufficiently large.} \tag{54}$$

Therefore,

$$P_e|_{\mathcal{W}=1} \;\; \leq \;\; \epsilon + \sum_{w=2}^{2^{NR}} \Pr(\tilde{E}_w), \tag{55}$$

where $\tilde{E}_w := E_w \wedge E_0^c \wedge E_1^c$.

Error event $\tilde{E}_w$: Let $\mathcal{L}_k$ be the set of nodes in layer $k$. We say that a node $j \in \mathcal{L}_k$ is *confused* [1] by $w$ if $(\underline{y}^{'}_{j}, w) \in \mathcal{T}^\epsilon$ for some $w \neq 1$. The destination is confused by $w$ under $\tilde{E}_w$. The source is not confused by definition. Hence, under the error event $\tilde{E}_w$, the nodes in the network get separated into two sets, ones that are confused by $w$, and others that are not confused by $w$. This is a cut in the network; see Figure 3 for an example.

Fix an arbitrary cut $\Omega$, and define the sets (see Figure 4)

$$F_{k,\Omega} := \mathcal{L}_k \cap \Omega \text{ and } G_{k,\Omega} := \mathcal{L}_k \cap \Omega^c. \tag{56}$$

[1] This is similar to the notion of *distinguishability* used in [5] and [2].

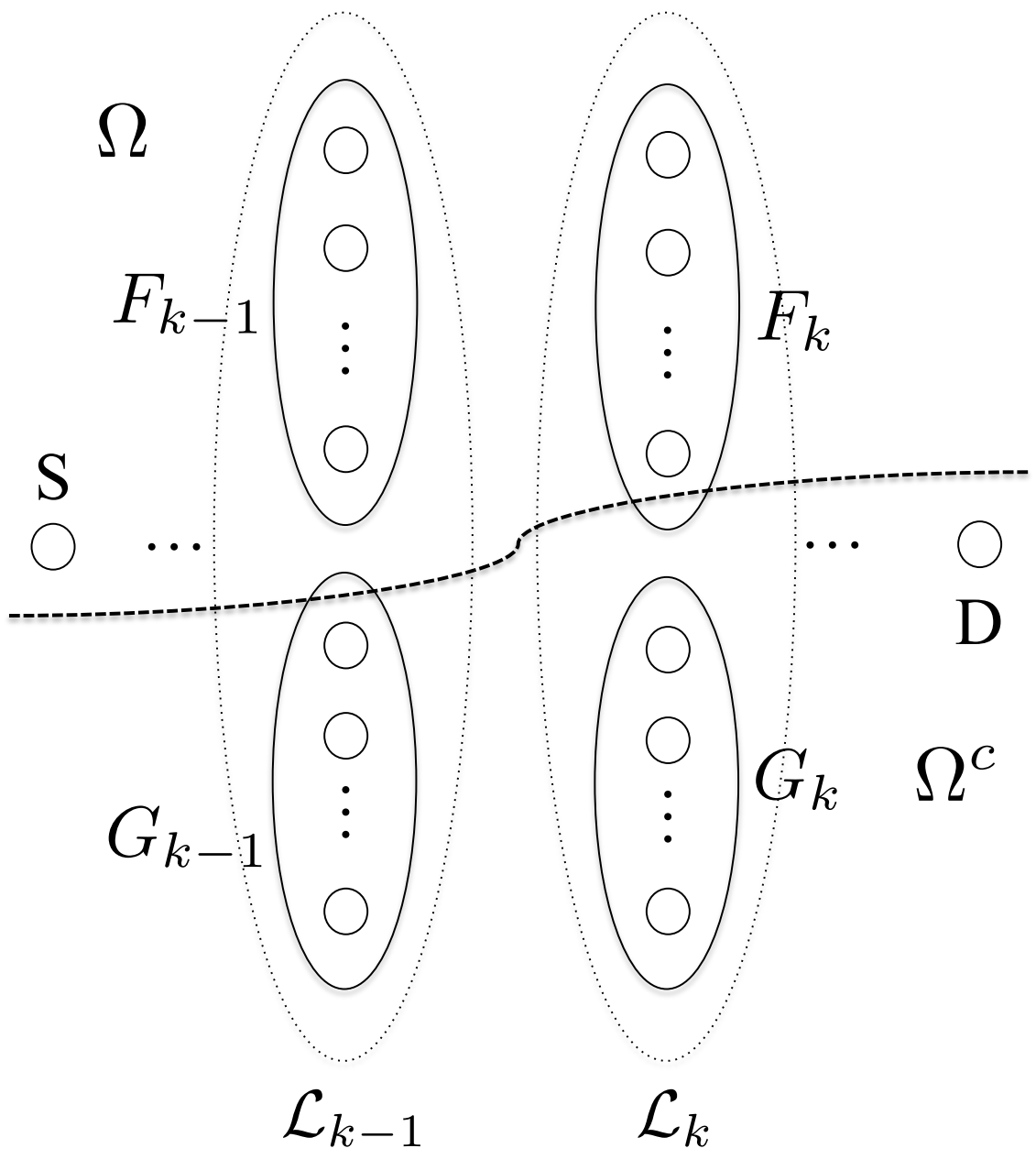

Fig. 4. Layers $(k-1)$ and $k$ in a network. The wireless links connecting the nodes are not shown. Each oval represents a particular set. The dashed curve in the center divides the network into $\Omega$ and $\Omega^c$.

Let us denote the concatenation of the received $2nN$-vectors of nodes at layer $\ell$ by $\underline{y}^{'}_{\mathcal{L}_\ell}$. Or, in greater detail, if $i_1, i_2, \ldots, i_q$ are the nodes, in lexicographic order, that are present in layer $\ell$, then $\underline{y}^{'}_{\mathcal{L}_\ell} := (\underline{y}^{'}_{i_1}, \underline{y}^{'}_{i_2}, \ldots, \underline{y}^{'}_{i_q})$. Correspondingly, we also define the concatenation of transmitted vectors at layer $\ell$ under $\underline{y}^{'}_{\mathcal{L}_\ell}$ by $\underline{x}_{\mathcal{L}_\ell}(\underline{y}^{'}_{\mathcal{L}_\ell}) := (A_{i_1}(\underline{y}^{'}_{i_1}), A_{i_2}(\underline{y}^{'}_{i_2}), \ldots, A_{i_q}(\underline{y}^{'}_{i_q}))$. Thereby we define the concatenated received vectors $\underline{y}^{'}_{\mathcal{L}_1}, \underline{y}^{'}_{\mathcal{L}_2}, \ldots, \underline{y}^{'}_{\mathcal{L}_L}$ at the $L$ layers in the network, and also the corresponding concatenated transmitted vectors. Next we define the *network-wide received vector* $\underline{y}^{'}_{\mathcal{V}} := (\underline{y}^{'}_{\mathcal{L}_1}, \underline{y}^{'}_{\mathcal{L}_2}, \ldots, \underline{y}^{'}_{\mathcal{L}_L})$, with receptions ordered according to layers, and lexicographically within layers. The corresponding *network-wide transmitted vector* is $\underline{x}_{\mathcal{V}}(\underline{y}^{'}_{\mathcal{V}}) := (\underline{x}_{\mathcal{L}_1}(\underline{y}^{'}_{\mathcal{L}_1}), \underline{x}_{\mathcal{L}_2}(\underline{y}^{'}_{\mathcal{L}_2}), \ldots, \underline{x}_{\mathcal{L}_L}(\underline{y}^{'}_{\mathcal{L}_L}))$.

We now wish to define the subset of network-wide received vectors that are consistent in the sense of joint typicality with a source message 1. Similar to the definition of $\mathcal{Y}^{'}_j(w)$ in (47) and (49), we define *received vectors at a node $j$ that are associated with message $w$,* except that we now do it at a layer and across the whole network. We can do this by exploiting the layered nature of the network where nodes at one layer only transmit to the nodes in the next layer.

We say *a network-wide received vector* $\underline{y}_{\mathcal{V}} = (\underline{y}'_{\mathcal{L}_1}, \underline{y}'_{\mathcal{L}_2}, \dots, \underline{y}'_{\mathcal{L}_L})$ *is associated with message* $w$ if for each $\ell$ and each node $j \in \mathcal{L}_\ell$, $(\underline{y}'_j, \underline{x}_{\mathcal{L}_{\ell-1}}(\underline{y}'_{\mathcal{L}_{\ell-1}})) \in \mathcal{T}_{\epsilon,p}$, where $p$ is the distribution $p(\tilde{x}_{\mathcal{L}_{\ell-1}}, \tilde{y}'_j)$, where $\tilde{x}_{\mathcal{L}_{\ell-1}}$ is uniformly distributed, and $p(\tilde{y}'_j|\tilde{x}_{\mathcal{L}_{\ell-1}})$ is the conditional distribution describing the channel from transmissions by nodes at layer $\ell - 1$ to node $j$. We denote by $\mathcal{Y}_{\mathcal{V}}(w)$ the set of such network-wide received vectors associated with message $w$, and denote by $\chi_{\mathcal{V}}(w)$ the set of network-wide transmitted vectors. If the source transmits message $w$, then with high probability, for a sufficiently large $N$, by the asymptotic equipartition property (AEP) of strongly typical vectors [20], the random network-wide vectors $\underline{y}_{\mathcal{V}}$ will lie in $\mathcal{Y}_{\mathcal{V}}(w)$.

We note that the codewords corresponding to the messages are random, as are the matrices $\{A_j\}$, due to the random coding strategy. As a consequence, $\mathcal{Y}_{\mathcal{V}}$ are also random, since they are dependent on the codewords and network coding matrices. Now we proceed to conduct an analysis under the random coding strategy.

Randomly choose a codeword for each message, and a set matrices $\{A_j : 1 \le j \le M-1\}$, and randomly, with a uniform distribution, pick a network-wide received vector $\underline{\tilde{y}}'_{\mathcal{V}}$ from the set $\mathcal{Y}_{\mathcal{V}}(1)$. For the random coding strategy, define $\mathcal{G}^w_{k,\Omega}$ as the event that every node in $G_{k,\Omega}$ is confused by $w$, i.e., for a node $j \in G_{k,\Omega}$,

$$(\{\underline{x}_i(w)\}_{i\in\mathcal{L}_{k-1}}, \underline{\tilde{y}}'_j) \in \mathcal{T}_{\epsilon,p}, \quad \text{for some } \underline{x}_i(w) \in \chi_i(w), \text{ for all } i \in \mathcal{L}_{k-1}, \tag{57}$$

with $p$ denoting the distribution $p(\{x_i\}_{i\in\mathcal{N}(j)}, \tilde{y}'_j)$, where $\{x_i\}_{i\in\mathcal{N}(j)}$ are independent and uniformly distributed and $p(\tilde{y}'_j|\{x_i\}_{i\in\mathcal{N}(j)})$ models the channel in the discrete network. Define $\mathcal{F}^w_{k,\Omega}$ as the event that no node in $F_{k,\Omega}$ is confused by $w$. Note that, in (57), due to the random coding strategy, $\chi_i(w)$ is also random.

For a random choice of $\underline{\tilde{y}}'_{\mathcal{V}}$, let $\mathcal{P}_{1,w,\Omega}$ denote the probability that no node in $\Omega$ is confused by $w$ when 1 is the transmitted message, and all the nodes in $\Omega^c$ are confused by $w$. Then,

$$\begin{aligned}
\mathcal{P}_{1,w,\Omega} &= \Pr(\bigwedge_{k\ge 0} \mathcal{F}^w_{k,\Omega} \wedge \mathcal{G}^w_{k,\Omega}) && (58)\\
&= \prod_{k\ge 1} \Pr(\mathcal{G}^w_{k,\Omega}| \bigwedge_{l=0}^{k-1} \mathcal{F}^w_{l,\Omega} \wedge \mathcal{G}^w_{l,\Omega}) \Pr(\mathcal{F}^w_{k,\Omega}|\mathcal{G}^w_{k,\Omega} \wedge \bigwedge_{l=0}^{k-1} \mathcal{F}^w_{l,\Omega} \wedge \mathcal{G}^w_{l,\Omega}) && (59)\\
&\le \prod_{k\ge 1} \Pr(\mathcal{G}^w_{k,\Omega}| \bigwedge_{l=0}^{k-1} \mathcal{F}^w_{l,\Omega} \wedge \mathcal{G}^w_{l,\Omega}). && (60)
\end{aligned}$$

Now we make two important observations concerning each relay node $j$'s buffering of $N$ of its received symbols and multiplication of such a buffered $2nN$-vector by the matrix $A_j$. First, since the matrix $A_j$ is independently chosen for each relay, all the relay mappings are independent of each other.

Second, when we randomize over all the linear encodings $A_j$ at relay node $j$, every transmit vector $\underline{x}_j \in \chi_j(w)$, for any $w$, is independently and uniformly distributed over the set of binary vectors of length $2nN$. Hence each symbol in a transmit vector $\underline{x}_j$ is independently and uniformly distributed.

Error event at layer $k$: Next, define $\mathcal{P}^k_{1,w,\Omega}$ as the contribution of the $k$-th layer to the probability, i.e.,

$$\mathcal{P}^k_{1,w,\Omega} \; := \; \Pr(\mathcal{G}^w_{k,\Omega} | \bigwedge_{l=0}^{k-1} \mathcal{F}^w_{l,\Omega} \wedge \mathcal{G}^w_{l,\Omega}). \tag{61}$$

Consider the transmit vectors of node $i \in \mathcal{L}_{k-1}$ given by $\underline{\tilde{x}}_i = A_i \underline{\tilde{y}}^{'}_i$. Under the conditioning events to determine $\mathcal{P}^k_{1,w,\Omega}$, we know that $\underline{\tilde{x}}_i \in \chi_i(w)$ for $i \in G_{k-1,\Omega}$, and $\underline{\tilde{x}}_i \in \chi_i(1)$ for $i \in F_{k-1,\Omega}$. Denote the transmit vector of node $i \in G_{k-1,\Omega}$ by $\underline{\tilde{x}}_i(w)$ to indicate that it is the transmit block of a confused node, and denote the transmit vector of node $i \in F_{k-1,\Omega}$ by $\underline{\tilde{x}}_i(1)$, since it is not confused by any other message. In order to compute $\mathcal{P}^k_{1,w,\Omega}$, we need to compute the probability that for every node $j \in G_{k,\Omega}$,

$$(\{\underline{x}_i(w)\}_{i \in \mathcal{L}_{k-1}}, \underline{\tilde{y}}_j) \in \mathcal{T}_{\epsilon,p}, \quad \text{for some } \underline{x}_i(w) \in \chi_i(w), \text{ for all } i \in \mathcal{L}_{k-1}, \tag{62}$$

where $p$ is the distribution $p(\tilde{x}_{\mathcal{L}_{k-1}}, \tilde{y}^{'}_j)$, where $\{x_i\}_{i \in \mathcal{L}_{k-1}}$ is uniformly distributed, and $p(\tilde{y}^{'}_j | \tilde{x}_{\mathcal{L}_{k-1}})$ is the conditional distribution describing the channel from transmissions by nodes at layer $k-1$ to node $j$, given that

- $\underline{\tilde{x}}_{F_{k-1,\Omega}}(1)$ was transmitted by the nodes in $F_{k-1,\Omega}$,
- $\underline{\tilde{x}}_{G_{k-1,\Omega}}(w)$ was transmitted by the nodes in $G_{k-1,\Omega}$, and
- $(\{\underline{\tilde{x}}_i(w)\}_{i \in G_{k-1,\Omega}}, \underline{\tilde{y}}_j) \in \mathcal{T}_{\epsilon,p}$, where $p$ is the distribution $p(\{x_i\}_{i \in G_{k-1,\Omega}}, \tilde{y}^{'}_j)$.

Project $\chi_{\mathcal{V}}(w)$, the set of network-wide transmit vectors, onto the nodes in $\mathcal{L}_{k-1}$ to obtain the set $\chi_{\mathcal{L}_{k-1}}(w)$. Pick $\mathcal{L}_{k-1}$-wide transmit vectors $\underline{x}_{\mathcal{L}_{k-1}} = \{\underline{x}_i(w)\}_{i \in \mathcal{L}_{k-1}}$ from $\chi_{\mathcal{L}_{k-1}}(w)$. If node $j$ checks to see if this choice of $x_{\mathcal{L}_{k-1}}$ will result in $(x_{\mathcal{L}_{k-1}}, \underline{\tilde{y}}^{'}_j)$ lying in $\mathcal{T}_{\epsilon,p}$, and $\underline{x}_i(w) = \underline{\tilde{x}}_i(w)$, for all $i \in G_{k-1,\Omega}$, then, under the conditions listed above, it will find a positive answer with

probability less than

$$2^{-NI(x_{F_{k-1,\Omega}};y_j^{'},x_{G_{k-1,\Omega}})} \;=\; 2^{-NI(x_{F_{k-1,\Omega}};y_j^{'}|x_{G_{k-1,\Omega}})}, \tag{63}$$

where the uniform distribution on $\{x_i\}$ is used to evaluate the mutual information and (63) is due to the independence of transmit symbols $\{x_i\}$. The choice of the i.i.d. uniform distribution for $\{x_i\}$ follows from the discussion following (60). Instead, for the set $\underline{x}_{\mathcal{L}_{k-1}}$, if $\underline{x}_i(w) \neq \underline{\tilde{x}}_i(w)$ for all $i \in G_{k-1,\Omega}$, then the probability of (62) is

$$2^{-NI(x_{F_{k-1,\Omega}},x_{G_{k-1,\Omega}};y_j^{'},)} \;\leq\; 2^{-NI(x_{F_{k-1,\Omega}};y_j^{'}|x_{G_{k-1,\Omega}})}. \tag{64}$$

For all other choices of $\{\underline{x}_i(w)\}_{i\in\mathcal{L}_{k-1}} \in \chi_{\mathcal{L}_{k-1}}(w)$, when if $\underline{x}_i(w) \neq \underline{\tilde{x}}_i(w)$ for node $i$ in a subset of $G_{k-1,\Omega}$, the probability of (62) is similarly upper bounded by

$$2^{-NI(x_{F_{k-1,\Omega}};y_j^{'}|x_{G_{k-1,\Omega}})}. \tag{65}$$

From Lemma C.1 (see Section C), $|\chi_{\mathcal{L}_{k-1}}(w)| = 2^{O(M)N}$. So, we apply the union bound with respect to all the vectors in $\chi_{\mathcal{L}_{k-1}}(w)$ to get

$$\begin{aligned} &\Pr(\text{node } j \text{ is confused by } w|\bigwedge_{l=0}^{k-1}\mathcal{F}_{l,\Omega}^{w}\wedge\mathcal{G}_{l,\Omega}^{w}) \\ &\leq\; 2^{O(M)N}\, 2^{-NI(x_{F_{k-1,\Omega}};y_j^{'}|x_{G_{k-1,\Omega}})}. \end{aligned} \tag{66}$$

We bound $\mathcal{P}_{1,w,\Omega}^{k}$ as

$$\begin{aligned} \mathcal{P}_{1,w,\Omega}^{k} &= \Pr(\text{nodes in } G_{k,\Omega} \text{ are confused by } w|\bigwedge_{l=0}^{k-1}\mathcal{F}_{l-1,\Omega}^{w}\wedge\mathcal{G}_{l-1,\Omega}^{w}) && (67)\\ &\leq \prod_{j\in G_{k,\Omega}}\Pr(\text{node } j \text{ is confused by } w|\bigwedge_{l=0}^{k-1}\mathcal{F}_{l-1,\Omega}^{w}\wedge\mathcal{G}_{l-1,\Omega}^{w}) && (68)\\ &\leq 2^{O(M)|G_{k,\Omega}|N}\, 2^{-N\sum_{j\in G_{k,\Omega}} I(x_{F_{k-1,\Omega}};y_j^{'}|x_{G_{k-1,\Omega}})} && (69)\\ &\leq 2^{O(M)|G_{k,\Omega}|N}\, 2^{-NI(x_{F_{k-1,\Omega}};y_{G_{k,\Omega}}^{'}|x_{G_{k-1,\Omega}})}. && (70) \end{aligned}$$

To obtain the bound in (68), we note that under the conditioning events, the nodes in $\mathcal{L}_{k-1}$ transmit $\underline{x}_{\mathcal{L}_{k-1}} = \{\underline{x}_i\}_{i\in\mathcal{L}_{k-1}}$. Given a set of transmissions, $\underline{x}_{\mathcal{L}_{k-1}}$, the received vectors $\underline{y}_{j_1}^{'}$ and $\underline{y}_{j_2}^{'}$ at node $j_1$ and $j_2$, respectively, in layer $k$ are conditionally independent. We obtain the bound by noting that the conditioning events in (68) involve many such mutually exclusive transmissions by the nodes in $\mathcal{L}_{k-1}$. The bound in (69) is from (66). With a lexicographic ordering of the nodes in $\mathcal{L}_k$ as $\{j_1, j_2, \ldots, j_{|G_{k,\Omega}|}\}$, the bound in (70) is obtained as

$$\begin{aligned}
&\sum_{l=1}^{|G_{k,\Omega}|} I(x_{F_{k-1,\Omega}}; y'_{j_l}|x_{G_{k-1,\Omega}}) \\
&= \sum_{l=1}^{|G_{k,\Omega}|} H(y'_{j_l}|x_{G_{k-1,\Omega}}) - H(y'_{j_l}|x_{\mathcal{L}_{k-1}}) && (71)\\
&\geq \sum_{l=1}^{|G_{k,\Omega}|} H(y'_{j_l}|x_{G_{k-1,\Omega}}, \cup_{m=1}^{l-1}\underline{y}'_{j_m}) - H(y'_{j_l}|x_{\mathcal{L}_{k-1}}, \cup_{m=1}^{l-1}\underline{y}'_{j_m}) && (72)\\
&= I(x_{F_{k-1,\Omega}}; y'_{G_{k,\Omega}}|x_{G_{k-1,\Omega}}), && (73)
\end{aligned}$$

where (72) follows from the Markov chain $(\cup_{m=1}^{l-1}\underline{y}'_{j_l}, x_{\mathcal{L}_{k-1}}) \to x_{\mathcal{L}_{k-1}} \to y'_{j_l}$, and that conditioning reduces entropy.

Back substitutions: Substituting from (70) in (60), we get

$$\begin{aligned}
\mathcal{P}_{1,w,\Omega} &\leq (2^{O(M)\sum_{k\geq 1}|G_{k,\Omega}|\,N})\,(2^{-N\sum_{k\geq 1} I(x_{F_{k-1,\Omega}}; y'_{G_{k,\Omega}}|x_{G_{k-1,\Omega}})}) && (74)\\
&\leq 2^{O(M)|\Omega^c|N}\, 2^{-NI(x_\Omega; y'_{\Omega^c}|x_{\Omega^c})} && (75)\\
&= 2^{O(M^2)N}\, 2^{-NI(x_\Omega; y'_{\Omega^c}|x_{\Omega^c})}. && (76)
\end{aligned}$$

The bound in (75) follows from the steps below and the chain rule for mutual information.

$$\begin{aligned}
I(x_{F_{k-1,\Omega}}; y'_{G_{k,\Omega}}|x_{G_{k-1,\Omega}}) &= H(y'_{G_{k,\Omega}}|x_{G_{k-1,\Omega}}) - H(y'_{G_{k,\Omega}}|x_{\mathcal{L}_{k-1}}) && (77)\\
&= H(y'_{G_{k,\Omega}}|x_{G_{k-1,\Omega}}) - H(y'_{G_{k,\Omega}}|x_{\mathcal{V}}, \cup_{\ell=1}^{k-1} y'_{G_{\ell,\Omega}}) && (78)\\
&\geq H(y'_{G_{k,\Omega}}|x_{\Omega^c}, \cup_{\ell=1}^{k-1} y'_{G_{\ell,\Omega}}) - H(y'_{G_{k,\Omega}}|x_{\mathcal{V}}, \cup_{\ell=1}^{k-1} y'_{G_{\ell,\Omega}}) && (79)\\
&\geq I(x_\Omega; y'_{G_{k,\Omega}}|x_{\Omega^c}, \cup_{\ell=1}^{k-1} y'_{G_{\ell,\Omega}}), && (80)
\end{aligned}$$

where the (78) uses the Markov structure of the layered network,

$$(x_{\mathcal{V}\setminus\mathcal{L}_{k-1}}, \cup_{\ell=1}^{k-1} y'_{G_{\ell,\Omega}}) \to x_{\mathcal{L}_{k-1}} \to y'_{G_{k,\Omega}}.$$

The probability of the event $\tilde{E}_w$ is bounded as

$$\begin{aligned}
\Pr(\tilde{E}_w) &\leq \sum_{\Omega\in\Lambda} \mathcal{P}_{1,w,\Omega} && (81)\\
&\leq \sum_{\Omega\in\Lambda} 2^{O(M^2)N}\, 2^{-NI(x_\Omega; y'_{\Omega^c}|x_{\Omega^c})} && (82)\\
&\leq 2^{M+1}\, 2^{O(M^2)N}\, 2^{-N\min_{\Omega\in\Lambda} I(x_\Omega; y'_{\Omega^c}|x_{\Omega^c})}. && (83)
\end{aligned}$$

Finally, substituting the bound for $\Pr(\tilde{E}_w)$ from (83) in (55) gives us

$$P_e|_{\mathcal{W}=1} \;\leq\; \epsilon + (2^{NR}-1)2^{M+1}\,2^{O(M^2)N}\,2^{-N\min_{\Omega\in\Lambda} I(x_\Omega; y'_{\Omega^c}|x_{\Omega^c})} \quad (84)$$

$$\leq\; \epsilon + 2^{M+1}2^{N(R+O(M^2)-\min_{\Omega\in\Lambda} I(x_\Omega; y'_{\Omega^c}|x_{\Omega^c}))}. \quad (85)$$

Hence, if $R < \min_{\Omega\in\Lambda} I(x_\Omega;\, y'_{\Omega^c}|x_{\Omega^c}) - O(M^2)$, then $P_e|_{\mathcal{W}=1}$ or $P_e$ can be made arbitrarily small for a sufficiently large $N$. ■

### *E. Approximate optimality of the linear network code for layered discrete networks*

Next we prove that the linear network code is approximately optimal for the layered network.

*Theorem 4.3: The linear network coding scheme achieves the capacity of the layered discrete network up to a bounded number of bits, i.e., the rate $R$ achieved by the coding scheme is bounded from the capacity $C_D$ of the layered discrete relay network by $O(M^2)$ bits,*

$$C_D - R \;=\; O(M^2), \quad (86)$$

*where the constant gap is independent of channel gains or SNR.*

*Proof:* In Lemma 3.1, we proved that the cut-set bound of the Gaussian network and its discrete counterpart are within a bounded gap of $O(M\log M)$ bits. In fact, in Section III-2 we also showed that $\overline{CS}_D = CS_G - O(M\log M)$, where $\overline{CS}_D$ is the value of the cut-set bound of the discrete network for the specific choice of the uniform i.i.d. input distribution (see (41)) and $CS_G$ is the cut-set bound of the Gaussian network. In Section III-1, we proved that $CS_G \geq CS_D$, where $CS_D$ is the cut-set bound of the discrete network. Therefore, $\overline{CS}_D = CS_D - O(M\log M)$. Since $CS_D \geq \overline{CS}_D$, we have

$$|CS_D - \overline{CS}_D| \;=\; O(M\log M). \quad (87)$$

From Lemma 4.2, we know that the linear network code achieves a rate $R$ within $O(M^2)$ bits of $\overline{CS}_D$. Hence, along with the bound in (87), the linear network code achieves rates within $O(M^2)$ bits of the cut-set bound $CS_D$. Since $CS_D \geq C_D$, the theorem is proved. ■

## V. LINEAR NETWORK CODE FOR LAYERED GAUSSIAN NETWORKS

In the earlier sections we proved that the cut-set bounds of the Gaussian and the discrete relay networks are within a bounded gap of $O(M\log M)$ bits. Later we developed a simple linear coding scheme for the layered discrete network that is approximately optimal. Also, the discrete model is a digital interface for operating the Gaussian networks; the signals in the discrete model are obtained by quantizing the signals in the Gaussian model. Combining these results we obtain a nearly capacity-achieving coding strategy for layered Gaussian relay networks, which consists of (i) quantizing received signals, and (ii) collecting a block of such signals and applying linear network coding on the overall vector of bits.

In this section, we prove the optimality of the linear network code for Gaussian relay networks and later extend this to MIMO Gaussian relay networks, where the nodes can have multiple transmit and receive antennas, and to multicast networks, where the source can transmit the same information to a subset of the nodes.

### A. Approximate optimality of the linear code for layered Gaussian relay networks

*Theorem 5.1: The linear network coding scheme achieves the capacity of the layered Gaussian relay network up to a bounded number of bits, i.e., the rate $R$ achieved by the linear network code is bounded from the capacity $C_G$ of the layered Gaussian relay network by $O(M^2)$ bits,*

$$C_G - R \;=\; O(M^2), \tag{88}$$

*where the constant gap is independent of channel gains or SNR.*

*Proof:* The linear network code is constructed over the layered discrete relay network as described in Section IV, and can be used on the Gaussian relay network as mentioned in Lemma 4.1.

From Theorem 4.3, we know that the linear network code achieves all rates within $O(M^2)$ bits of the cut-set bound of the layered discrete relay network $CS_D$. In Lemma 3.1, we proved that $|CS_G - CS_D| \;=\; O(M \log M)$. Combining these results, and noting that $CS_G \geq C_G$, we get the statement of the theorem. ■

The theorem establishes that the discrete model can serve as a digital interface for Gaussian networks, since a coding scheme for the layered discrete relay network involving simple linear operations is approximately optimal and can be used on the layered Gaussian relay network. Since the linear network code achieves rates within a bounded bit gap of $CS_G$, the theorem is also a proof of the near-optimality of the cut-set bound for layered Gaussian relay networks, though this was proved earlier in [2] for a smaller bounded gap of $O(M \log M)$ bits.

As a consequence of the above theorem, we can also prove the following lemma.

*Lemma 5.2: The capacity of the layered discrete relay network and the capacity of the layered Gaussian relay network are within a bounded gap of* $O(M^2)$ *bits, i.e.,* $|C_G - C_D| = O(M^2)$*. Furthermore, a near-optimal code for the layered discrete relay network is also a near-optimal code for the layered Gaussian network.*

*Proof:* The linear network code is near-optimal for both the layered discrete relay network and the original layered Gaussian relay network and achieves rates within $O(M^2)$ bits of the capacities of both the networks (see Theorems 4.3 and 5.1). Hence, it follows that the capacity of the layered Gaussian and the layered discrete network are within a bounded gap of $O(M^2)$ bits, i.e., $|C_G - C_D| = O(M^2)$.

Next consider a near-optimal code for the layered discrete network that achieves rates within a SNR-independent gap from the capacity of the discrete network. From Lemma 4.1, we can lift this coding strategy to obtain a code for the original Gaussian relay network without any decrease in the rate. Since the capacity of the two networks are within a bounded gap, the lifted code is a near-optimal code for the layered Gaussian relay network. ■

### *B. MIMO relay networks*

We can extend the linear network coding scheme to layered MIMO networks, where nodes have multiple transmit and receive antennas. In (2), we defined the received signal in a MIMO receiver in a Gaussian relay network. We operate the MIMO Gaussian relay network on the discrete interface as described in Section II-B. The linear network code is defined on the MIMO discrete relay network. The basic ideas in the coding scheme remain the same, but with some modifications to accommodate multiple antennas. The details are:

*1) Source's codewords:* The source constructs a set of $2^{NR}$ codewords of length $N$. Every codeword is a $T_0 \times N$ matrix, where each row of the matrix is transmitted by one of the transmit antennas of the source. Each entry in the matrix is a complex number from the QAM constellation (or equivalently, a $2n_{MIMO}$-length binary tuple, see Section II-B) that is independently chosen with the uniform distribution.

*2) Relay's mappings:* Relay $j$ has $U_j$ receive antennas, receives a vector of $U_j$ symbols every instant, and buffers $N$ such vectors. The received binary tuples are adjoined to construct a $2n_{MIMO}NU_j$-length binary vector. The relay constructs a $2n_{MIMO}NT_j \times 2n_{MIMO}NU_j$ binary matrix $A_{j,MIMO}$ where each entry of $A_{j,MIMO}$ is either $0$ or $1$. It multiplies the $2n_{MIMO}NU_j$-length binary vector with $A_{j,MIMO}$ to obtain a $2n_{MIMO}NT_j$-length binary vector. It splits this vector into $T_j$ vectors of length $2n_{MIMO}N$. Each tuple of length $2n_{MIMO}N$ corresponds to $N$ transmit symbols for a particular transmit antenna.

*3) Decoding at the destination:* Destination collects $N$ received signals and finds a message that is associated with the received vector; see the decoding procedure in Section IV-C3 for details. If it finds more than one message that satisfies this condition or if it finds none, it declares an error else it declares the unique transmitted message.

Let the maximum number of transmit or receive antennas in the network be $T_{\max}$.

*Theorem 5.3: The linear network coding scheme achieves the capacity of the layered MIMO Gaussian relay network up to a bounded number of bits, i.e., the linear code achieves all rates $R$ bounded from the capacity $C_{G,MIMO}$ of the layered Gaussian relay network as*

$$C_{G,MIMO} - R \;=\; O(M^2 T_{\max}). \tag{89}$$

*The constant in the bounded gap is independent of channel gains or SNR.*

*Proof:* The steps in the proof are essentially the same as that of Theorem 5.1:

1) Operate the MIMO Gaussian relay network on the digital interface defined by the discrete model.
2) First we prove that the cut-set bound on the capacity of the MIMO Gaussian relay network and the MIMO discrete network are within a bounded gap of $O(MT_{\max} \log MT_{\max})$ bits. Here, the proof of Lemma 4.2 can be reused by a simple observation. While evaluating the cut-set bound, we can view each transmit and receive antenna as a virtual node. Hence the total number of nodes in the network is at most $MT_{\max}$ which gives us the necessary bound.
3) The coding scheme achieves all rates within $O(M^2 T_{\max})$ bits of the cut-set bound of the discrete MIMO network evaluated for a specific choice of the input distribution. Here the input distribution is i.i.d. across all the transmit antennas in the network and is uniform over the alphabet. The arguments in the proof of Lemma 4.2 carry over to MIMO networks by noting that due to multiple transmit and receive antennas, the receive and transmit signals are complex vectors instead of scalars. In the proof, the value of $|\chi_i(w)|$ for MIMO networks will be different than before. For MIMO networks, the bound on $H(\underline{y}'_{\mathcal{L}_2}|\underline{x}_{\mathcal{L}_1})$ in (116) increases to $T_{\max}|\mathcal{L}_1|\,O(N)$. Due to this, the bound on $|\Psi(w)|$ is $2^{O(MT_{\max})N}$. With this, we get the required bounded gap of $O(M^2T_{\max})$ from the cut-set bound.

4) Then it is straightforward to prove near-optimality of the linear code for MIMO Gaussian networks; see Theorem 5.1.

■

We state the counterpart of Lemma 5.2 for MIMO relay networks.

*Lemma 5.4: The capacity of the layered MIMO discrete relay network and the capacity of the layered MIMO Gaussian relay network is within a bounded gap of $O(M^2T_{\max})$ bits, i.e., $|C_{G,MIMO} - C_{D,MIMO}| = O(M^2T_{\max})$. Furthermore, a near-optimal code for the layered MIMO discrete relay network is also a near-optimal code for the layered MIMO Gaussian network.*

*Proof:* Same as the proof of Lemma 5.2. ■

### C. Multicast networks

In a multicast network, the source node wants to communicate the same information to a subset of the nodes (instead of a single destination as in the previous sections). The remaining nodes which are not the intended recipients act as relays. Let $\mathcal{D}$ be the set of nodes that are the intended recipients of the source's message. Then the cut-set bound on the capacity $C_{G,Mult}$ for such networks is given by

$$C_{G,Mult} \;\leq\; \max_{p(x_0,x_1,\ldots,x_{M-1})}\; \min_{D\in\mathcal{D}}\; \min_{\Omega\,\in\,\Lambda_D}\; I(x_\Omega; y_{\Omega^c}|x_{\Omega^c}), \tag{90}$$

where $\Lambda_D$ is the set of all cuts in the network that separate the source from the destination $D$. We can extend the linear network code from Section IV-C to this class of networks, with the only difference being that *all* the intended destinations in $\mathcal{D}$ decode the source's transmission. In a layered multicast network, the destinations in $\mathcal{D}$ can be spread across the various layers in the network.

*Theorem 5.5: The linear network coding scheme achieves the capacity of the multicast layered Gaussian network up to a bounded number of bits, i.e., the linear network code achieves any rate $R$ which is bounded from the capacity $C_{G,Mult}$ of the multicast layered Gaussian network as*

$$C_{G,Mult} - R \;=\; O(M^2). \tag{91}$$

*The constant in the bounded gap is independent of channel gains or SNR.*

*Proof:* The proof of this theorem resembles that of Theorem 5.1. We give a general outline of the proof and skip the details.

1) Operate the multicast Gaussian network on the digital interface defined by the discrete model.
2) First we prove that the cut-set bounds on the capacity of the multicast Gaussian network and the multicast discrete network are within a bounded gap of $O(M \log M)$ bits. Though the cut-set bound for multicast networks is slightly different from that of relay networks, the proof of Lemma 4.2 can be reused by individually comparing the cut-set bound between the source and each destination in $\mathcal{D}$.
3) Then we prove that the coding scheme achieves all rates within $O(M^2)$ bits of the cut-set bound of the discrete network, when the cut-set is evaluated for i.i.d. uniform inputs. We re-use the proof of Lemma 4.2, with the only difference being that multiple destinations want to decode the source's transmission instead of a single destination.
4) With the above arguments, we can prove the equivalent of Theorem 4.3 for layered multicast networks. Then, with the same arguments as in the proof of Theorem 5.1, we can prove the near-optimality of the linear code up to $O(M^2)$ bits for the original layered multicast Gaussian network.

■

We state the counterpart of Lemma 5.2 for multicast relay networks.

*Lemma 5.6: The capacity of the layered multicast discrete relay network and the capacity of the layered multicast Gaussian relay network is within a bounded gap of $O(M^2)$ bits, i.e., $|C_{G,Mult} - C_{D,Mult}| = O(M^2)$. Furthermore, a near-optimal code for the layered multicast discrete relay network is also a near-optimal code for the layered multicast Gaussian network.*

*Proof:* Same as the proof of Lemma 5.2. ■

## VI. Concluding remarks

Our overall near capacity achieving strategy for a layered Gaussian relay network can be summarized as follows. The number $n$ of bits of precision is chosen as the logarithm of the largest real or imaginary part of any channel gain. Then by simple quantization and truncation, we create a purely discrete but stochastic network. This layered discrete network is operated in an en bloc fashion. At each node received signals are buffered for a block, based on which, transmit signals in the next block are generated. The coding strategy is particularly simple. Linear network coding is performed by simply multiplying the buffered vector by a square random binary matrix. The resulting long vector is broken into symbols for use in the next block. A similar strategy can be used for MIMO nodes, and also for multicast relay networks.

The above strategy has the advantage that it employs a simple coding strategy requiring minimal signal processing at the relays. We note that most codes for relay networks do require considerable signal processing by the relay nodes consisting of non-linear operations such as vector quantization, compression, decoding, etc. In the linear coding strategy presented in this paper, each relay performs scalar quantization followed by a simple matrix multiplication. Non-linearity is thus introduced into the code due to quantization. Introducing some non-linearity is unavoidable in the strategy, indeed necessary, due to the inability of linear codes to achieve the capacity within a constant gap in the Gaussian relay network.

The linear network code is a robust scheme in the sense that the relay need not know the channel gains on either the incoming or the outgoing links. Since the transmit and receive signals are quantized to binary tuples of length $2n$, all the nodes only need to know the global parameter $n$. The quantization requirements of the linear network code are completely defined by the parameter $n$, which therefore also determines the resolution of the analog-to-digital convertor (ADC) and digital-to-analog convertor (DAC) that are required for operating the network within a bounded gap from the network capacity.

The random matrix at the relays interleaves all the bits and perhaps increases the complexity of the decoding algorithm at the destination. It is an interesting problem to simplify this by constructing explicit encoding matrices for the relays that preserve the properties of a random matrix. This might help us construct a graphical model for describing the channel from the source to the destination and lead to a low-complexity code for relay networks that is decodable with iterative message passing algorithms.

For a general Gaussian network with many sources, many destinations, relays, and any data transmission requirements, one can similarly construct the corresponding discrete counterpart and prove that the capacity region of the corresponding discrete network is contained in the capacity region of the Gaussian network. The converse remains to be proved however, if true:

that the capacity region of the original Gaussian network is contained in that of the discrete network, up to a constant gap.

Since the network coding scheme proposed here for the wireless relay network case is the same as the network coding scheme that achieves the capacity of the wireline and the linear deterministic network, it is possible that the body of results developed in network coding could be possibly used to solve other problems in wireless networks too.

## APPENDIX A
## LOSS OF CAPACITY FOR UNIFORM INPUTS

*Lemma A.1:* Consider a cut $\Omega$ in a Gaussian network. Then the loss in the mutual information for choice of fractional inputs (as defined in Step 2 of Section III-2) is $O(M)$.

*Proof:* Denote the i.i.d. Gaussian inputs by $\{x_i^G\}$. Also denote i.i.d Gaussian inputs distributed as $\mathcal{CN}(0, 1/3)$ by $\{\tilde{x}_i^G\}$. Now consider the fractional inputs defined in Section III-2, denoted by $\{x_i^F\}$.

We observe some similarities between $\{\tilde{x}_i^G\}$ and $\{x_i^F\}$. Both the sets of inputs are i.i.d. and each input has the same variance in both cases. Hence, the covariance matrix of $\underline{\tilde{x}}^G = (\tilde{x}_i^G)$ or $\underline{x}^F = (x_i^F)$ is the scaled identity matrix $I_{M\times M}/3$. Denote the received signal in $\Omega^c$ under the Gaussian inputs and the fractional inputs by $\tilde{y}_{\Omega^c}^G$ and $y_{\Omega^c}^F$, respectively. It follows that the covariance matrices of the vectors $(\underline{\tilde{x}}^G, \tilde{y}_{\Omega^c}^G)$ and $(\underline{x}^F, y_{\Omega^c}^F)$ are the same.

Now consider the mutual information $I(x_\Omega^F; y_{\Omega^c}^F | x_{\Omega^c}^F)$,

$$\begin{aligned}
I(x_\Omega^F; y_{\Omega^c}^F | x_{\Omega^c}^F) &= h(x_\Omega^F | x_{\Omega^c}^F) - h(x_\Omega^F | x_{\Omega^c}^F, y_{\Omega^c}^F) && (92)\\
&= \sum_{i\in\Omega} h(x_i^F) - h(\underline{x}_\Omega^F | \underline{x}_{\Omega^c}^F, \underline{y}_{\Omega^c}^F) && (93)\\
&= \sum_{i\in\Omega} (2\log(1/\sqrt{2})) - h(x_\Omega^F | x_{\Omega^c}^F, y_{\Omega^c}^F) && (94)\\
&= -|\Omega|\log 2 - h(x_\Omega^F | x_{\Omega^c}^F, y_{\Omega^c}^F). && (95)
\end{aligned}$$

where (93) follows from the independence of $\{x_i\}$, and (94) follows from direct computation of the differential entropy. Now, for the circular Gaussian inputs

$$h(\tilde{x}_\Omega^G | \tilde{x}_{\Omega^c}^G) = \sum_{i\in\Omega} h(\tilde{x}_i^G) = \log(\pi e/3)|\Omega|. \qquad (96)$$

Since, for a given covariance constraint, the conditional entropy is maximized by the Gaussian distribution with the same covariance [22],

$$h(x^F_\Omega|x^F_{\Omega^c}, y^F_{\Omega^c}) \;\leq\; h(\tilde{x}^G_\Omega|\tilde{x}^G_{\Omega^c}, \tilde{y}^G_{\Omega^c}). \tag{97}$$

Substituting (96) and (97) into (95),

$$\begin{aligned} I(x^F_\Omega; y^F_{\Omega^c}|x^F_{\Omega^c}) &\geq h(\tilde{x}^G_\Omega|\tilde{x}^G_{\Omega^c}) - h(\tilde{x}^G_\Omega|\tilde{x}^G_{\Omega^c}, y^G_{\Omega^c}) - |\Omega|(\log(\pi e/3) + \log 2) & (98)\\ &\geq I(\tilde{x}^G_\Omega; \tilde{y}^G_{\Omega^c}|\tilde{x}^G_{\Omega^c}) - O(M). & (99)\end{aligned}$$

Now the mutual information with i.i.d. $\mathcal{CN}(0,1)$ inputs $\{x^G_i\}$ is given by

$$I(x^G_\Omega; y^G_{\Omega^c}|x^G_{\Omega^c}) \;=\; \log|I + \mathcal{H}_\Omega \mathcal{H}^\dagger_\Omega|, \tag{100}$$

where $\mathcal{H}_\Omega$ is the channel transfer matrix across the cut $\Omega$. Now consider the same cut, but increase the noise variance at all the received signals from $1$ to $3$. The mutual information reduces to $\log|I + \mathcal{H}_\Omega \mathcal{H}^\dagger_\Omega/3|$. Since the effect of increasing the noise variance is the same as that of reducing the signal power to $1/3$,

$$I(\tilde{x}^G_\Omega; \tilde{y}^G_{\Omega^c}|\tilde{x}^G_{\Omega^c}) \;=\; \log|I + \mathcal{H}_\Omega \mathcal{H}^\dagger_\Omega/3|. \tag{101}$$

Comparing (100) and (101),

$$I(\tilde{x}^G_\Omega; \tilde{y}^G_{\Omega^c}|\tilde{x}^G_{\Omega^c}) \;\geq\; I(x^G_\Omega; y^G_{\Omega^c}|x^G_{\Omega^c}) - 3|\Omega|. \tag{102}$$

Hence

$$I(x^F_\Omega; y^F_{\Omega^c}|x^F_{\Omega^c}) \;\geq\; I(x^G_\Omega; y^G_{\Omega^c}|x^G_{\Omega^c}) - O(M). \tag{103}$$

In [2], it is shown that the loss in choosing the inputs to be i.i.d. Gaussian $\mathcal{CN}(0,1)$ instead of the joint Gaussian distribution is $O(M)$ bits. Therefore the choice of fractional inputs $\{x^F_i\}$ leads to a loss of at most $O(M)$ bits in the mutual information when compared to the joint Gaussian inputs. ■

## APPENDIX B
## PROBABILITY OF THE ERROR EVENT $E_1$

*Lemma B.1:* The probability of the events $E_0$ or $E_1$: $(\underline{y}_M, 1) \notin \mathcal{T}$, for any $\epsilon > 0$, is bounded by

$$\Pr(E_0) + \Pr(E_1) \;\leq\; \epsilon, \text{ for } N \text{ sufficiently large.} \tag{104}$$

*Proof:* The proof of this lemma involves repeated application of the AEP for strongly typical vectors (see [20]).

Suppose the source transmits the codeword $\underline{x}_0(1)$. From the AEP of strongly typical vectors, the received signal $\underline{y}_j^{'}$ for a node in layer 1 is jointly typical with $\underline{x}_0(1)$, with probability exceeding $1 - \epsilon_1$, for a sufficiently large $N$, for any positive $\epsilon_1$. Hence, with high probability, $(\underline{y}_j^{'}, w) \in \mathcal{T}$ for $j \in \mathcal{L}_1$. Node $i$ in $\mathcal{L}_1$ subsequently transmits a message vector $\underline{x}_i = A_i \underline{y}_i^{'}$, with $\underline{x}_i \in \chi_i(1)$. Then, the received vector $\underline{y}_j^{'}$ at a node in $\mathcal{L}_2$ satisfies (by AEP of strongly typical vectors)

$$(\underline{y}_j^{'}, \{\underline{x}_i\}_{i \in \mathcal{N}(j)}) \in \mathcal{T}_{\epsilon,p}, \tag{105}$$

with probability exceeding $1-\epsilon_2$, for a sufficiently large $N$, for any positive $\epsilon_2$. In (105), the joint typicality of the vectors in $\mathcal{T}_\epsilon$ is with respect to the joint distribution $p(\{x_i\}_{i \in \mathcal{N}(j)}, y_j^{'})$, where $\{x_i\}_{i \in \mathcal{N}(j)}$ are independent and uniformly distributed, and $p(y_j^{'} | \{x_i\}_{i \in \mathcal{N}(j)})$ models the channel in the discrete network. Hence, with high probability, $(\underline{y}_j^{'}, 1) \in \mathcal{T}$ for $j \in \mathcal{L}_2$. Subsequently, each node in $\mathcal{L}_2$ transmits a message vector from $\chi_j(1)$.

We carry out this analysis across all the layers in the network, and obtain that the received signal in $\mathcal{L}_k$ satisfies $(\underline{y}_j^{'}, 1) \in \mathcal{T}$ with probability exceeding $1-\epsilon_k$, for a sufficiently large $N$, for any $\epsilon_k > 0$. The error event $E_1$ occurs if the received vector at the destination is not associated with message 1, and this will not occur (with high probability) if all the received vectors at the intermediate relay nodes are associated with 1. Hence, we get

$$\Pr(E_1) \;\leq\; \sum_k \epsilon_k, \text{ for a sufficiently large } N, \tag{106}$$

where the summation is over the number of layers.

Similarly to the above arguments, we can prove that the transmit and received message vectors in the network are strongly typical, with probability exceeding $1 - \epsilon_0$, for a sufficiently large $N$, for any $\epsilon_0 > 0$. Hence

$$\Pr(E_0) \;\leq\; \epsilon_0. \tag{107}$$

Since $\{\epsilon_k\}$ are arbitrary positive numbers, for any $\epsilon > 0$,

$$\Pr(E_0) + \Pr(E_1) \;\leq\; \epsilon, \text{ for a sufficiently large } N. \tag{108}$$

■

# APPENDIX C

## BOUNDING THE SIZE OF THE TYPICAL SET

*Lemma C.1:* The size of the set $|\chi_{\mathcal{L}_{k-1}}(w)|$ (with the notations as given in the proof of Lemma 4.2) is $2^{O(M)\,N}$, where the size is independent of channel gains or SNR.

*Proof:* Consider the set $\mathcal{Y}_{\mathcal{V}}(w)$ of network-wide received vectors associated with the message $w$, defined in the proof of Lemma 4.2. For every $\underline{y}_{\mathcal{V}} \in \mathcal{Y}_{\mathcal{V}}(w)$, we can define a network-wide transmitted vector, associated with the message $w$, as

$$\underline{x}_{\mathcal{V}}(\underline{y}'_{\mathcal{V}}) := (\underline{x}_{\mathcal{L}_1}(\underline{y}'_{\mathcal{L}_1}), \underline{x}_{\mathcal{L}_2}(\underline{y}'_{\mathcal{L}_2}), \dots, \underline{x}_{\mathcal{L}_L}(\underline{y}'_{\mathcal{L}_L})). \tag{109}$$

Define $\chi_{\mathcal{V}}(w)$ to consist of all such network-wide transmitted vectors associated with message $w$.

Consider the relay nodes in $\mathcal{L}_1$ and let us bound the conditional entropy $H((y'_i)_{i\in\mathcal{L}_1}|x_0)$:

$$H(y'_{\mathcal{L}_1}|x_0) \leq \sum_{i\in\mathcal{L}_1} H(y'_i|\underline{x}_0) \tag{110}$$

$$= \sum_{i\in\mathcal{L}_1} H([h_{1i}x_0 + z_i]|x_0) \tag{111}$$

$$\leq \sum_{i\in\mathcal{L}_1} H(z'_i, c_i), \tag{112}$$

where $z'_{jn}$ is the integer part of the Gaussian noise[2] and $c_{jn}$ is the carry from adding the fractional parts of the signal and the noise. Since $z_j$ is distributed as $\mathcal{CN}(0,1)$, $H(z'_j)$ is independent of channel gains or SNR and $H(\underline{z}'_j) = O(1)$. Also $H(c_j) = O(1)$, hence $H(y'_{\mathcal{L}_1}|x_0) = |\mathcal{L}_1|O(1)$.

Let $\mathcal{Y}_{\mathcal{L}_1}(w)$ be the projection of $\mathcal{Y}_{\mathcal{V}}(w)$ onto the nodes in $\mathcal{L}_1$. For a sufficiently large $N$, by the AEP of strongly typical vectors, the size of $\mathcal{Y}_{\mathcal{L}_1}(w)$ is

$$|\mathcal{Y}_{\mathcal{L}_1}(w)| = 2^{NH(y'_{\mathcal{L}_1}|x_0)} = 2^{N|\mathcal{L}_1|O(1)}. \tag{113}$$

Since every set of transmit vectors in $\chi_{\mathcal{L}_1}(w)$ is associated with a received vector in $\mathcal{Y}_{\mathcal{L}_1}(w)$, the size of $\chi_{\mathcal{L}_1}(w)$ is also bounded by

$$|\chi_{\mathcal{L}_1}(w)| \leq 2^{N|\mathcal{L}_1|O(1)}. \tag{114}$$

[2] $z_{jn}$ is a complex number lying in $\mathbb{Z} + \imath\mathbb{Z}$ corresponding to the integer portion of the real and imaginary parts of the complex noise $z_j$.

Now, similarly to (110)–(112), we can bound

$$H(y'_{\mathcal{L}_2}|x_{\mathcal{L}_1}) \;\leq\; \sum_{j\in\mathcal{L}_2} H(y'_j|x_{\mathcal{L}_1}) \tag{115}$$

$$= \; |\mathcal{L}_2|O(1). \tag{116}$$

Let us fix a set of transmit vectors $\underline{x}_{\mathcal{L}_1} \in \chi_{\mathcal{L}_1}(w)$; then consider a set of received vectors $\underline{y}'_{\mathcal{L}_2}$ for the nodes in $\mathcal{L}_2$ such that

$$(\underline{y}'_j, \{\underline{x}_i\}_{i\in\mathcal{N}(j)}) \in \mathcal{T}_{\epsilon,p}, \text{ for all } j \in \mathcal{L}_2. \tag{117}$$

The distribution $p$ is $p(\{\tilde{x}_i\}_{i\in\mathcal{N}_j}, \tilde{y}'_j)$, where $\{\tilde{x}_i\}_{i\in\mathcal{N}(j)}$ are independent and uniformly distributed, and $p(\tilde{y}'_j|\{\tilde{x}_i\}_{i\in\mathcal{N}(j)})$ models the channel from the nodes in $\mathcal{N}(j)$ to $j$ in the discrete network. For a fixed $\underline{x}_{\mathcal{L}_1}$, the size of the set of received vectors $y'_{\mathcal{L}_2}$ that satisfy the above is given by (for a sufficiently large $N$, by the joint AEP of strongly typical sequences)

$$2^{NH(y'_{\mathcal{L}_2}|x_{\mathcal{L}_1})} \;\leq\; 2^{N\mathcal{L}_2 O(1)}. \tag{118}$$

Define $\mathcal{Y}_{\mathcal{L}_1,\mathcal{L}_2}(w)$ as the projection of $\mathcal{Y}_{\mathcal{V}}(w)$ onto the nodes in $\mathcal{L}_1$ and $\mathcal{L}_2$. Now every set of vectors in $\mathcal{Y}_{\mathcal{L}_1,\mathcal{L}_2}(w)$ can be obtained by first fixing a set of received vectors $\underline{y}'_{\mathcal{L}_1}$ in $\mathcal{Y}_{\mathcal{L}_1}(w)$, determining the transmit vectors generated from it as $\underline{x}_{\mathcal{L}_1}(\underline{y}'_{\mathcal{L}_1})$, and finding a set of received vectors $\underline{y}'_{\mathcal{L}_2}$ that are jointly typical with this set $\underline{x}_{\mathcal{L}_1}(\underline{y}'_{\mathcal{L}_1})$, as in (117). By counting the number of vectors across the layers, the size of $\mathcal{Y}_{\mathcal{L}_1,\mathcal{L}_2}(w)$ is bounded by

$$|\mathcal{Y}_{\mathcal{L}_1,\mathcal{L}_2}(w)| \;\leq\; (2^{N\mathcal{L}_1 O(1)})(2^{N\mathcal{L}_2 O(1)}) \;=\; 2^{N(\mathcal{L}_1+\mathcal{L}_2)O(1)}. \tag{119}$$

With the bound, obtained similarly to (112),

$$H(y'_{\mathcal{L}_k}|x_{\mathcal{L}_{k-1}}) \;\leq\; |\mathcal{L}_k|O(1), \tag{120}$$

we extend this argument across all the layers in the network. The size of $\mathcal{Y}_{\mathcal{V}}(w)$ can be bounded as

$$|\mathcal{Y}_{\mathcal{V}}(w)| \;\leq\; 2^{N\sum_{\ell=1}^{L}|\mathcal{L}_\ell|O(1)} \;=\; 2^{NO(M)}. \tag{121}$$

Since each network-wide transmit vector in $\chi_{\mathcal{V}}(w)$ is a function of a vector in $\mathcal{Y}_{\mathcal{V}}(w)$,

$$|\chi_{\mathcal{V}}(w)| \;=\; 2^{NO(M)}. \tag{122}$$

Since $\chi_{\mathcal{L}_{k-1}}(w)$ is the projection of the set $\chi_{\mathcal{V}}(w)$ onto the nodes in $\mathcal{L}_{k-1}$, the size of $\chi_{\mathcal{L}_{k-1}}(w)$ is bounded by the size of $\chi_{\mathcal{V}}(w)$. Hence, the lemma is proved. ■